\documentclass[conference,pdftex]{IEEEtran}
\usepackage{graphicx}
\usepackage[nobreak]{cite}
\usepackage{amsmath,amssymb}
\usepackage{newtxmath}
\usepackage{framed}
\usepackage{xspace}
\newcommand{\Heading}[1]{\subsubsection{#1}}
\newcommand{\RQ}[1]{\textit{RQ}${}_{\mathrm{#1}}$}
\newcommand{\Conclusion}[1]{\begin{framed}\noindent #1\end{framed}}

\makeatletter
\let\MYcaption\@makecaption
\makeatother
\usepackage[font=footnotesize, subrefformat=parens]{subcaption}
\makeatletter
\let\@makecaption\MYcaption
\makeatother

\newcommand{\etal}{\textit{et al.}\xspace}
\newcommand{\figref}[1]{Fig.~\ref{#1}}
\newcommand{\Figref}[1]{Figure~\ref{#1}}

\newcommand{\Id}[1]{\texttt{#1}}
\newcommand{\Word}[1]{``\textit{#1}''}
\newcommand{\Rename}[2]{\Id{#1} $\to$ \Id{#2}}
\newcommand{\Intent}[1]{\textsc{#1}}
\newcommand{\Insert}[1]{\Intent{Insert}(\Word{#1})}
\newcommand{\Delete}[1]{\Intent{Delete}(\Word{#1})}
\newcommand{\Replace}[2]{\Intent{Replace}(\Word{#1}, \Word{#2})}
\newcommand{\Other}[1]{\Intent{Other}(\Word{#1})}
\newcommand{\Inflect}[1]{\Intent{Inflect}(\Word{#1})}

\newcommand{\Rel}[1]{\text{\textsc{#1}}}
\newcommand{\BelongsC}{$\Rel{Belongs}_C$\xspace}
\newcommand{\BelongsM}{$\Rel{Belongs}_M$\xspace}
\newcommand{\BelongsF}{$\Rel{Belongs}_F$\xspace}
\newcommand{\BelongsA}{$\Rel{Belongs}_A$\xspace}
\newcommand{\BelongsL}{$\Rel{Belongs}_L$\xspace}
\newcommand{\CoOccursM}{$\Rel{Co-occurs}_M$\xspace}
\newcommand{\Extends}{$\Rel{Extends}$\xspace}
\newcommand{\Implements}{$\Rel{Implements}$\xspace}
\newcommand{\TypeM}{$\Rel{Type}_M$\xspace}
\newcommand{\TypeV}{$\Rel{Type}_V$\xspace}
\newcommand{\Invokes}{$\Rel{Invokes}$\xspace}
\newcommand{\Accesses}{$\Rel{Accesses}$\xspace}
\newcommand{\Passes}{$\Rel{Passes}$\xspace}
\newcommand{\Assigns}{$\Rel{Assigns}$\xspace}

\begin{document}

\title{Empirical Study of Co-Renamed Identifiers}

\author{\IEEEauthorblockN{Yuki Osumi, Naotaka Umekawa, Hitomi Komata, and Shinpei Hayashi}
\IEEEauthorblockA{
\textit{Tokyo Institute of Technology}, 
Tokyo 152--8550, Japan \\
osumi@se.c.titech.ac.jp, hayashi@c.titech.ac.jp}
}

\maketitle

\thispagestyle{plain}

\begin{abstract}
\emph{Background:}
The renaming of program identifiers is the most common refactoring operation.
Because some identifiers are related to each other, developers may need to rename related identifiers together.
\emph{Aims:}
To understand how developers rename multiple identifiers simultaneously, it is necessary to consider the relationships between identifiers in the program and the brief matching for non-identical but semantically similar identifiers.
\emph{Method:}
We investigate the relationships between co-renamed identifiers and identify the types of their relationships that contribute to improving the recommendation using more than 1M of renaming instances collected from the histories of open-source software projects.
We also evaluate and compare the impact of co-renaming and the relationships between identifiers when inflections occur in the words in identifiers are taken into account.
\emph{Results:}
We revealed several relationships of identifiers that are frequently found in the co-renamed identifiers, such as the identifiers of \emph{methods in the same class} or \emph{an identifier defining a variable and another used for initializing the variable}, depending on the type of the renamed identifiers.
Additionally, the consideration of inflections did not affect the tendency of the relationships.
\emph{Conclusion:}
These results suggest an approach that prioritizes the identifiers to be recommended depending on their types and the type of the renamed identifier.
\end{abstract}

\begin{IEEEkeywords}
identifier, rename refactoring, refactoring recommendation
\end{IEEEkeywords}

\section{Introduction}\label{s:introduction}
Program identifiers play a significant role in program comprehension, and identifiers are frequently renamed during software refactoring.
It is estimated that identifiers account for approximately 70\% of the total program content \cite{deissenboeck2006concise}.
Corazza \etal also reported that developers can facilitate knowledge transfer among developers by naming identifiers appropriately that reflect their intentions and the domain knowledge \cite{LINSEN-corazza-icsm2012}.
Developers can infer the intent and the behavior of code fragments if the identifiers in them have appropriate names \cite{RefactoringWorkbook}.
Therefore, appropriately naming identifiers and/or renaming them can make it easier for developers to guess the roles of identifiers and improve the understanding of the program.
Renaming identifiers is considered as the most common form of refactoring \cite{murphy2006java}.

When an identifier is renamed, it may be necessary to rename other identifiers related to the renamed identifier.
This is because there may be a misunderstanding if multiple expressions refer to the same concept to be used for naming identifiers \cite{deissenboeck2006concise}.
However, it is time-consuming to identify all the identifiers that need to be renamed.
The replacement feature of source code editors and rename refactoring tools provided by current integrated development environments~(IDEs) cannot determine all the identifiers that need to be renamed together.
Therefore, developers must determine the identifiers to be renamed manually, which is difficult and may lead to missing changes.

In this study, we investigated co-renamed program identifiers to discuss their critical characteristics for improving the recommendation of identifiers to be co-renamed.
We surveyed co-renamed identifiers extracted from past histories of open-source software~(OSS) repositories, clarified the relationships among them that were prone to be correlated, and discussed the relationships that should be prioritized in the recommendation.
In addition, we evaluated the variation of word stemming in identifiers and discussed the influence of inflections, i.e., the influence of singular and plural and those of participles, on the recommendation.
There are several attempts to utilize the relationship among identifiers for renaming recommendations \cite{Expanding-Conducted-Rename-Refactorings, Thies}.
The results of our empirical study can strengthen their approaches by prioritizing the relationships to be used for the renaming recommendation.

The main contributions of this work are summarized as follows:
\begin{itemize}
  \item a large-scale empirical study using OSS repositories on the relationship between co-renamed identifiers and
  \item results of evaluating the effect of inflections in identifiers on co-renaming.
\end{itemize}
As a result of evaluating co-renamed identifiers, we identified the relationships among identifiers that frequently occur in renamed identifiers.
Additionally, the frequency of relationships differed depending on the type of renamed identifiers.
Furthermore, we found that the consideration of inflections did not affect the tendency of the relationships.

The rest of this paper is organized as follows.
In the next section, we explain our aim to support identifier co-renaming via an example of co-renamed identifiers.
Section~\ref{s:technique} presents our evaluation methods.
Section~\ref{s:result} explains our evaluation results and discusses their significance for renaming recommendation.
In Section~\ref{s:relatedwork}, we summarize existing work on identifiers and their renaming.
Finally, Section~\ref{s:conclusion} concludes this paper and summarizes future work.

\section{Motivation}\label{s:motivation}

\begin{figure}[tb]\centering
  \begin{minipage}{\hsize}\centering
    \includegraphics[width=\linewidth]{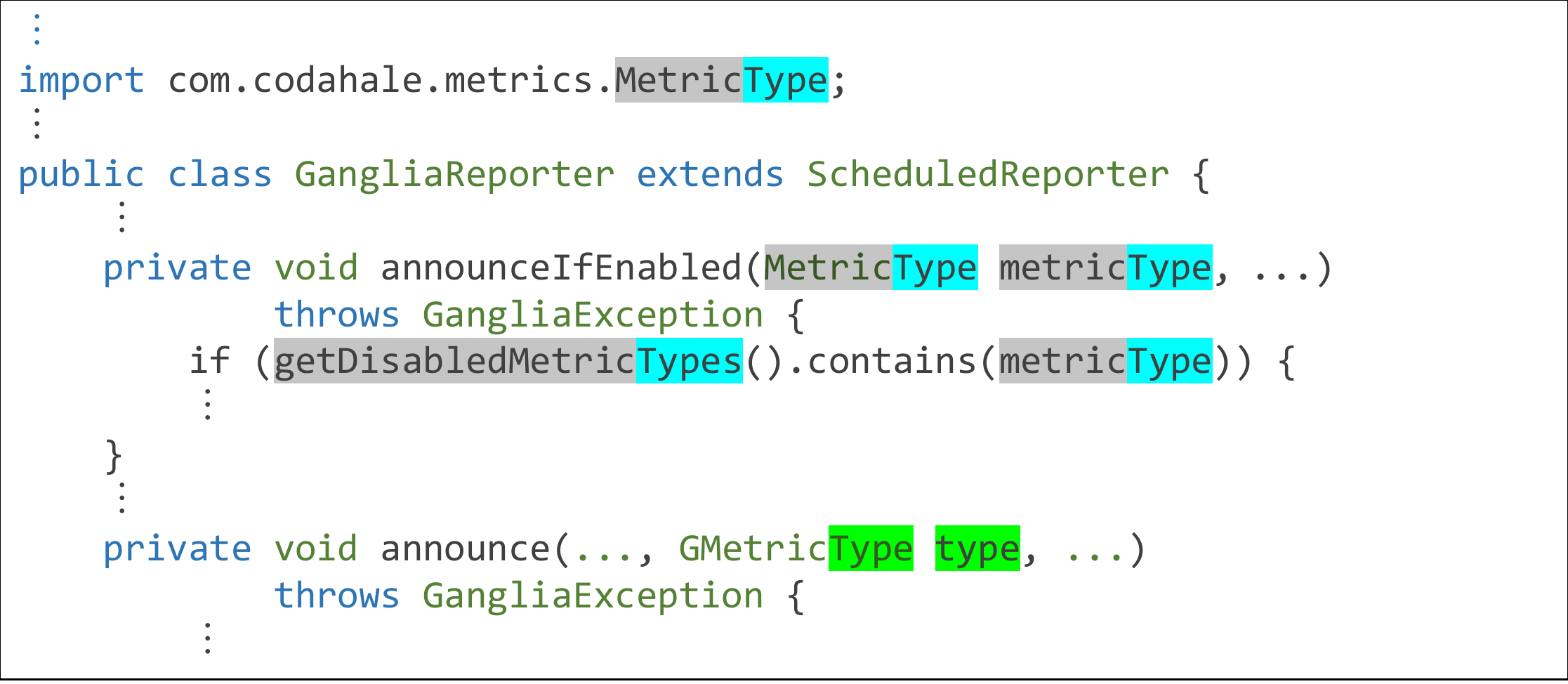}
    \subcaption{Before renaming.}\label{f:norm_before}
  \end{minipage} \\ \vspace{1.5em} 
  \begin{minipage}{\hsize}\centering
    \includegraphics[width=\linewidth]{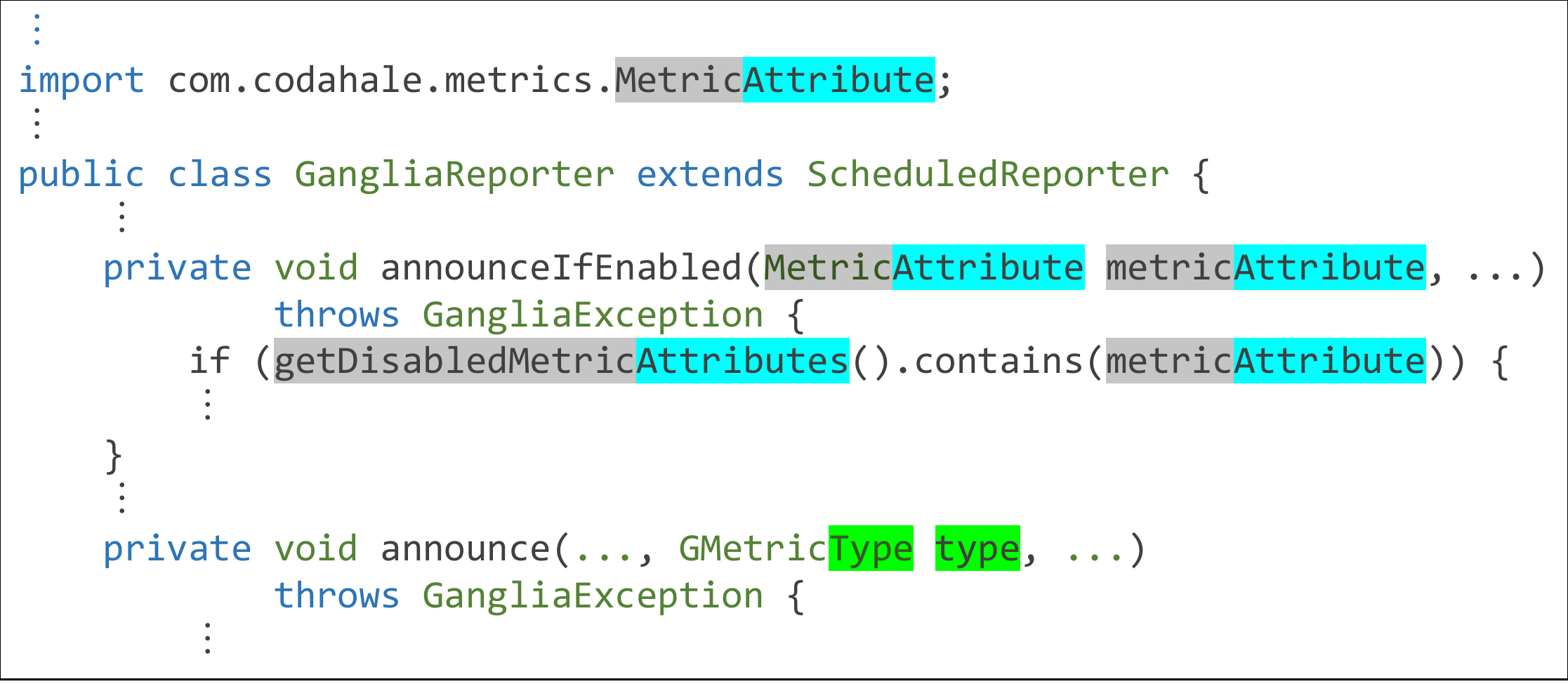}
    \subcaption{After renaming.}\label{f:norm_after}
  \end{minipage}
  \caption{Renamed identifiers in Metric project.}\label{f:motivation}
\end{figure}
When an identifier is renamed, another identifier may need to be renamed accordingly.
This is because the existence of multiple representations of the same concept may cause developers to misunderstand the program \cite{deissenboeck2006concise}.
An example of renamed identifiers is shown in \figref{f:motivation}\footnote{https://github.com/dropwizard/metrics/commit/3ccd7a1}.
In this commit, the class \Id{MetricType} is renamed to \Id{MetricAttribute}.
From the renaming, we can infer that the developer intended to replace the word \Word{type} in the identifiers with \Word{attribute}.
The gray-highlighted identifiers in the figure are those renamed in this commit.
The blue-highlighted parts in the figure represent the changes based on the developer's intention, i.e., \Word{type} before the renaming to \Word{attribute} after the renaming.
In contrast, the green parts in the figure indicate that they were not renamed; the word \Word{type} remained after this co-renaming operations.

However, it is time-consuming for developers to determine all the identifiers that should be renamed.
For example, although the source code editors' \emph{Replace All} feature can replace all the matched strings simultaneously, it may lead to errors as it applies to elements that need not be replaced.
Additionally, rename refactoring tools provided by IDEs such as Eclipse\footnote{https://www.eclipse.org/} and IntelliJ IDEA\footnote{https://www.jetbrains.com/idea/} can limit the identifiers to be renamed at once to those of the same program element; developers cannot change multiple identifiers to be renamed at once.
Therefore, developers need to identify all the identifiers to be renamed simultaneously, which is a challenging task wasting much time.

According to the renamed one, recommending possible candidates of related identifiers to be renamed together can assist developers in improving the consistency of their source code.
When a developer renames the class \Id{MetricType} to \Id{MetricAttribute} as shown in \figref{f:motivation}, by applying the substitution from the word \Word{type} to \Word{attribute} to the variable \Id{metricType}, which is typed as \Id{MetricType}, we can recommend a renaming to \Id{metricAttribute}.

In recommending identifiers to be renamed, the following should be taken into account:
\begin{itemize}
  \item relationships between identifiers that are likely to be renamed together and
  \item inflections in identifiers.
\end{itemize}

Because not all identifiers should be renamed together to follow the developer's renaming intention, it is useful to consider the relationship between identifiers that are likely to be co-renamed.
For example, on one hand, the substitution in \figref{f:motivation} from the word \Word{type} to \Word{attribute} in the class \Id{MetricType} is also applicable to the word \Word{type} in the class \Id{GMetricType}, but this did not happen; \Id{GMetricType} was not renamed.
On the other hand, this substitution was applied to the variable \Id{metricType}, whose type is \Id{MetricType}, and the variable identifier was renamed to \Id{metricAttribute}.
Here, a relationship of ``variable and its type'' exists between these two identifiers: the class \Id{MetricType} and the variable \Id{metricType}.
It is useful to clarify the kinds of relationships between the identifiers that are likely to be renamed simultaneously.

Additionally, if inflections are not taken into account, the developer's renaming intention cannot be applied, and recommendations may fail.
For example, in the renaming in \figref{f:motivation}, the method \Id{getDisabledMetricTypes} has also been renamed, and a similar substitution from the word \Word{type} to \Word{attribute} has occurred; however, the plural form \Word{types} has been replaced with \Word{attributes}, unlike the case of \Id{MetricType}.
These renamings should be considered as those based on the same intention.

To clarify the characteristics of the relationships between identifiers to be co-renamed together and the effects of inflections in identifiers, we conducted an empirical study of co-renamed identifiers.
Through answers to the following research questions~(RQs), we discuss the aspects that should be considered in recommending identifiers to be co-renamed.
\def\RQone{How often do co-renamings occur?}
\def\RQtwo{To what extent do co-renamed identifiers correlate the relationships of identifiers?}
\def\RQthree{What is the difference when the inflections in identifiers are taken into account?}
\begin{itemize}
  \item \textbf{\RQ{1}}: \RQone
  \item \textbf{\RQ{2}}: \RQtwo
  \item \textbf{\RQ{3}}: \RQthree
\end{itemize}
In \RQ{1}, we confirm that a large number of co-renamed identifiers occur.
In \RQ{2}, we characterize the relationships between identifiers that have undergone co-renamings.
In \RQ{3}, we clarify the effects of inflections in identifiers on the number of co-renamings and the relationships between identifiers that have undergone co-renamings.

\section{Methodology}\label{s:technique}

\begin{figure}[tb]\centering
  \includegraphics[width=8.8cm]{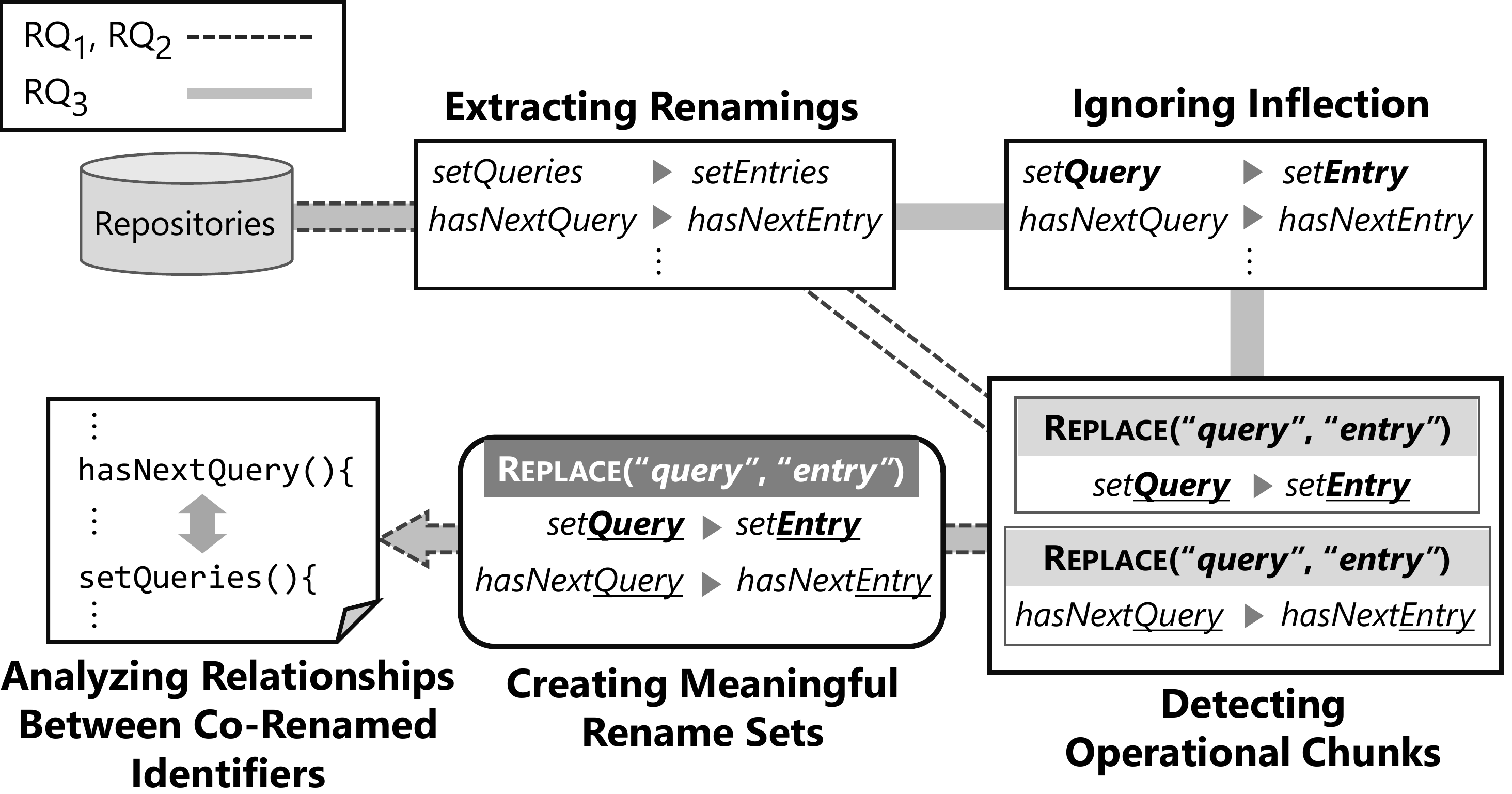}
  \caption{Overview of our study.}\label{f:research_process}
\end{figure}

\subsection{Overview}\label{s:all}
\Figref{f:research_process} shows the overview of our study.
First, we extract all the renamings from the commit history of the version control repositories using an existing refactoring detection tool.
Next, we lemmatize each word in the identifier names before and after each extracted renaming to ignore the inflection.
Subsequently, we compare the old identifier name with the new one to obtain the operational chunks of the renaming.
We then create meaningful rename sets that share their operational chunks.
Finally, we examine the relationship between their target identifier names in the source codes for each pair of two renamings in a meaningful rename set.
Note that we skip the third step when answering \RQ{1} and \RQ{2} since we consider the inflection only when answering \RQ{3}.

\subsection{Extracting Renamings}\label{s:extract}
We detected renamings using RefactoringMiner ver.\ 2.0.2 \cite{tsantalis2018accurate,tsantalis2020refactoringminer} in 187 of the repositories used by Silva \etal and Tsantalis \etal in their refactoring studies \cite{why-we-refactor,tsantalis2018accurate}.
For each repository, we extracted class renamings~(\textsf{Rename Class}), method renamings (\textsf{Rename Method}), class attribute renamings (\textsf{Rename Attribute}), method parameter renamings (\textsf{Rename Parameter}), and variable renamings~(\textsf{Rename Variable}) from detected refactorings and created a set $\mathbb{R}$ of all the renamings.
Let $r.\mathit{commit}$ denote the commit in which a renaming $r\in \mathbb{R}$ was performed. 
We excluded repositories that did not contain any renamings, resulting in 176 repositories with a history from June 2001 to January 2021.
The total number of commits in these repositories was 1,883,276, whereas the total number of extracted renamings was 1,084,121.

\subsection{Ignoring Inflection}\label{s:norm}
We split identifier names into words using Ronin in the Spiral \cite{hucka2018spiral} package for each identifier name before and after a renaming.
For each word, we converted it to lower case and lemmatized it to ignore inflection using WordNetLemmatizer \cite{bird2009natural}.

\subsection{Detecting Operational Chunks}\label{s:detect}
Using a difference extraction algorithm based on the longest continuous matching subsequence detection, we extracted each consecutive addition, deletion, and replacement of words as one \emph{operational chunk} from the identifier names before and after a renaming $r$.
We denote the all set of operational chunks of $r$ by $r.\mathit{chunks}$.
For each renaming $r$, we split identifier names before and after the renaming $r$ into words using Ronin and created word sequences.
We detected $r.\mathit{chunks}$ from the differences between the word sequence before and after renaming $r$.
An operational chunk consists of the sequence of words changed by renaming and changing types of words. 
By introducing operational chunks, we can regard renamings for different identifier names as a co-renaming that shares the same operational chunk.

There are four types of operational chunks as listed below.
\begin{itemize}
  \item \Intent{Insert}(\textit{added words}): Addition of words. 
  For example, a renaming \Rename{data\-Provider\-Id}{data\-Provider\-\underline{Instance}\-Id} involves an operational chunk \Insert{instance}.
  \item \Intent{Delete}(\textit{deleted words}): Deletion of words.
  For example, a renaming \Rename{Skip\-\underline{Constant}\-Result}{Skip\-Result} involves an operational chunk \Delete{Constant}.
  \item \Intent{Replace}(\textit{deleted words}, \textit{added words}): Replace of words.
  For example, a renaming \Rename{\underline{get}\-Random}{\underline{create}\-Random} involves an operational chunk \Replace{get}{create}.
  \item \Intent{Other}(\textit{words}): Other type of renaming.
  For example, a renaming \Rename{TIMES}{times} downcases its word, which belongs to an operational chunk \Other{time}.
\end{itemize}
It should be noted that \Intent{Other} is only detected when none of \Intent{Insert}, \Intent{Delete}, or \Intent{Replace} is detected in a given renaming.
When detecting \Intent{Other}, we compared the identifier names before and after a renaming, starting from the first word, and detected \Intent{Other} for each word that differs.

\subsection{Creating Meaningful Rename Sets}\label{s:create}
Based on the operational chunks detected in Section~\ref{s:detect}, we created a meaningful rename set, a collection of renamings that share the same operational chunk in the same commit.
We defined a meaningful rename set $U_{c,h}$ in commit $c$, where its belonging operational chunk is $h$, as follows:
\[
  U_{c,h} \coloneq \{r\in\mathbb{R} \mid r.\mathit{commit} = c\ \wedge\ r.\mathit{chunks} \supseteq h \}.
\]
For the set $\mathbb{R}$ of all the renamings in each repository, we created a set $\mathbb{U}$ of meaningful rename sets, i.e., $\mathbb{U}=\{U_{c,h} \mid U_{c,h}\neq\emptyset \}$.

When we detected multiple operational chunks in a renaming, we created meaningful rename sets for each operational chunk and included the renaming in all their sets.
For example, renaming $r$: \Rename{minimumVersion}{versionSpec} is an element of both $U_\text{\Delete{minimum}}$ and $U_\text{\Insert{spec}}$.

\subsection{Analyzing Relationships between Co-Renamed Identifiers}\label{s:analyze}
For each pair of two renamings in a meaningful rename set $U_{c,h}$, we evaluated the relationship between the renamed identifiers in the source codes.
Let $P(U_{c,h})$ denote a set of all pairs of two renamings in $U_{c,h}$, i.e.,
\[
  P(U_{c,h}) \coloneq \{(r_i, r_j) \mid r_i,r_j\in U_{c,h}\ \wedge\ r_i\neq r_j \}.
\]
For a pair $p_i= (r_i, r_j)\in P(U_{c,h})$, we detected relationships $R_c(p_i)$ from the source codes of the parent commit of commit $c$, i.e., the source codes before $r_i$ and $r_j$ perform as follows:
\[
  R_c(p_i) = \mathit{relation}_c(r_i.\mathit{old},\ r_j.\mathit{old})
\]
where $r.\mathit{old}$ is the identifier before $r$ is performed and $\mathit{relation}_c(I_i, I_j)$ is a set of relationships between identifiers $I_i$ and $I_j$ in the source codes of the parent commit of $c$.
We defined 14 relationships between co-renamed identifiers and categorized them into three: Location, Type, and Call/Data Dependency.
The details of each relationship \textbf{R} are as follows.

\noindent 1. Location
\begin{itemize}
  \item \BelongsC: Relationship between a class $c$ and its inner class $c.c$.
  \item \BelongsM: Relationship between a class $c$ and its method $c.m$.
  \item \BelongsF: Relationship between a class $c$ and its attribute $c.a$.
  \item \BelongsA: Relationship between a method $m$ and its parameter $a$.
  \item \BelongsL: Relationship between a method $m$ and its local variable $v$.
  \item \CoOccursM: Relationship between two methods $c.m_1$ and $c.m_2$ in the same class $c$.
\end{itemize}

\noindent 2. Type
\begin{itemize}
  \item \Extends: Relationship between a class $c$ and its direct subclass $c'$.
  \item \Implements: Relationship between an interface $i$ and the class $c$ that implements $i$.
  \item \TypeM: Relationship between a method $m$ and its return type $c$.
  \item \TypeV: Relationship between an attribute, a variable, or a method parameter $v$ and its type $c$.
\end{itemize}

\noindent 3. Call and Data Dependency
\begin{itemize}
  \item \Invokes: Relationship between a method $m_1$ and another method $m_2$ invoked by $m_1$ ($m_1 \neq m_2$).
  \item \Accesses: Relationship between a method $c.m$ of a class $c$ and the attribute $c.a$ of the same class $c$ referenced by $c.m$.
  \item \Assigns: Relationship between a left side attribute or variable $v_{\mathit{left}}$ of an assignment statement and a right side attribute of the assignment statement $v_{\mathit{right}}$ such as an attribute, a parameter, a variable or an invocation of a method.
  \item \Passes: Relationship between a (formal) parameter $p$ of a method $m$ and an argument (actual parameter) $a$ of the method $m$ such as an attribute, a variable, or invocation of a method.
\end{itemize}

See again \figref{f:motivation}(a) with assuming that the developer triggered to rename the class name \Id{MetricType} to \Id{MetricAttribute}.
In this code fragment, the type of the parameter \Id{metricType} is \Id{MetricType}, which means that there is a relationship of \TypeV between the parameter \Id{metricType} and the class \Id{MetricType}.
Also, the method \Id{getDisabledMetricTypes} returns an object typed as \verb|Set<MetricType>|, which is not expressed in \figref{f:motivation}(a) though, which leads to a relationship of \TypeM between the method \Id{getDisabledMetricTypes} and the class \Id{MetricType}.
We identify these kinds of relationships by applying simple static analyses for parsed source files.

We converted the target source codes to XML files using srcML \cite{collard2004document} and detected $\mathit{relation}_c(r_i.\mathit{old}, r_j.\mathit{old})$ using XPath.
For each relationship \textbf{R}, we defined XPaths that return nodes only if a pair $(r_i.\mathit{old}, r_j.\mathit{old})$ satisfies \textbf{R}.
For example, in \BelongsM, we defined an XPath that returns nodes only if a class $c$ has a method $m$ as follows.
\begin{itemize}
  \item //class[name[text()=$c$]]/block/function[name[text()=$m$]]
\end{itemize}
For a pair $(r_i.\mathit{old}, r_j.\mathit{old})$, we created a set of relationships \textbf{R} that the pair satisfied and regarded the set as a $\mathit{relation}_c(r_i.\mathit{old}, r_j.\mathit{old})$.

\section{Empirical Study}\label{s:result}
\subsection{\RQ{1}: \RQone}

\Heading{Motivation}
The purpose of this RQ is to confirm that many renamings co-occur.
We determined the number of co-renamed identifiers and the characteristics of these renamings.

\Heading{Study Design}
For each of the 176 repositories in Section~\ref{s:extract}, we evaluated a percentage of co-renamings.
For each $\mathbb{U}$ created in each repository, we computed the ratio of the total number of elements of meaningful rename sets that contained more than one element $\sum_{U\in \mathbb{U}, |U|\geq 2} |U|$ to the total number of all elements of meaningful rename sets $\sum_{U\in \mathbb{U}} |U|$.
We regarded this ratio as the rate of co-renamings.

We determined the ratio of the number of co-renamed identifiers to the number of all the co-renamed identifiers for each co-renamed identifier.
For the case where $n$ identifiers are co-renamed, we computed its ratio of the total number $\sum_{U\in \mathbb{U}, |U| = n} |U|$ to all the co-renamings $\sum_{U\in \mathbb{U}, |U| \geq 2} |U|$.

For each number of co-renamed identifiers, we also determined the ratio of the number of unique identifier names included in the co-renamings to the number of co-renamed identifiers.
Because some co-renames contain renamings for the same identifier names, the number of renamed identifier names in the co-renamings may be small even when the number of co-renamed identifiers is large.
We computed the rate of the total number of meaningful rename sets in which $m$ unique identifier names were involved to the number of meaningful rename sets whose size was $n$.
Here, the number of unique identifier names $m$ was computed as the number of non-overlapping pre-renaming identifier names $|\{r.\mathit{old} \mid r\in U\}|$ in a meaningful rename set.

\begin{figure}[tb]\centering
      \includegraphics[scale=0.35]{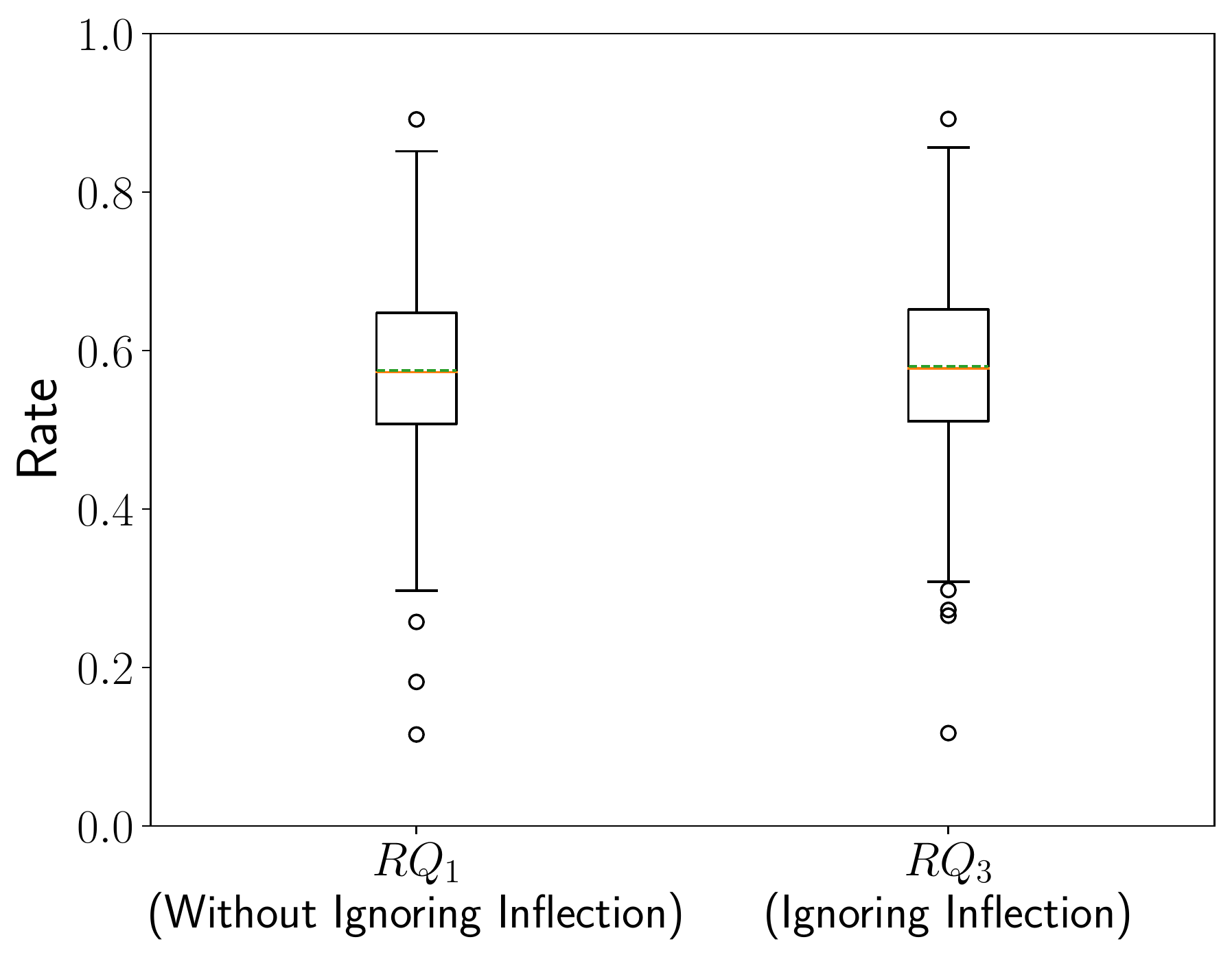}
      \caption{Rate of co-renamings.}\label{f:rq1_rq3_co_rename}
\end{figure}

\Heading{Results}
On average, 57\% of the renamings were co-renamings.
The left side of \figref{f:rq1_rq3_co_rename} is a box plot showing the ratio of the total number of elements in the meaningful rename sets with more than one element to the total number of elements in all the meaningful rename sets, i.e., the rate of co-renamings in each repository.
The vertical axis represents the rate of co-renamings.
Although the rates are widely distributed (16\% to 89\%), the first quartile is 51\%, which shows that 75\% of all projects contain more than 50\% of co-renamings.
The rate of co-renamings is 0.57 for both the mean and median.

\begin{figure}[tb]\centering
  \includegraphics[scale=0.35]{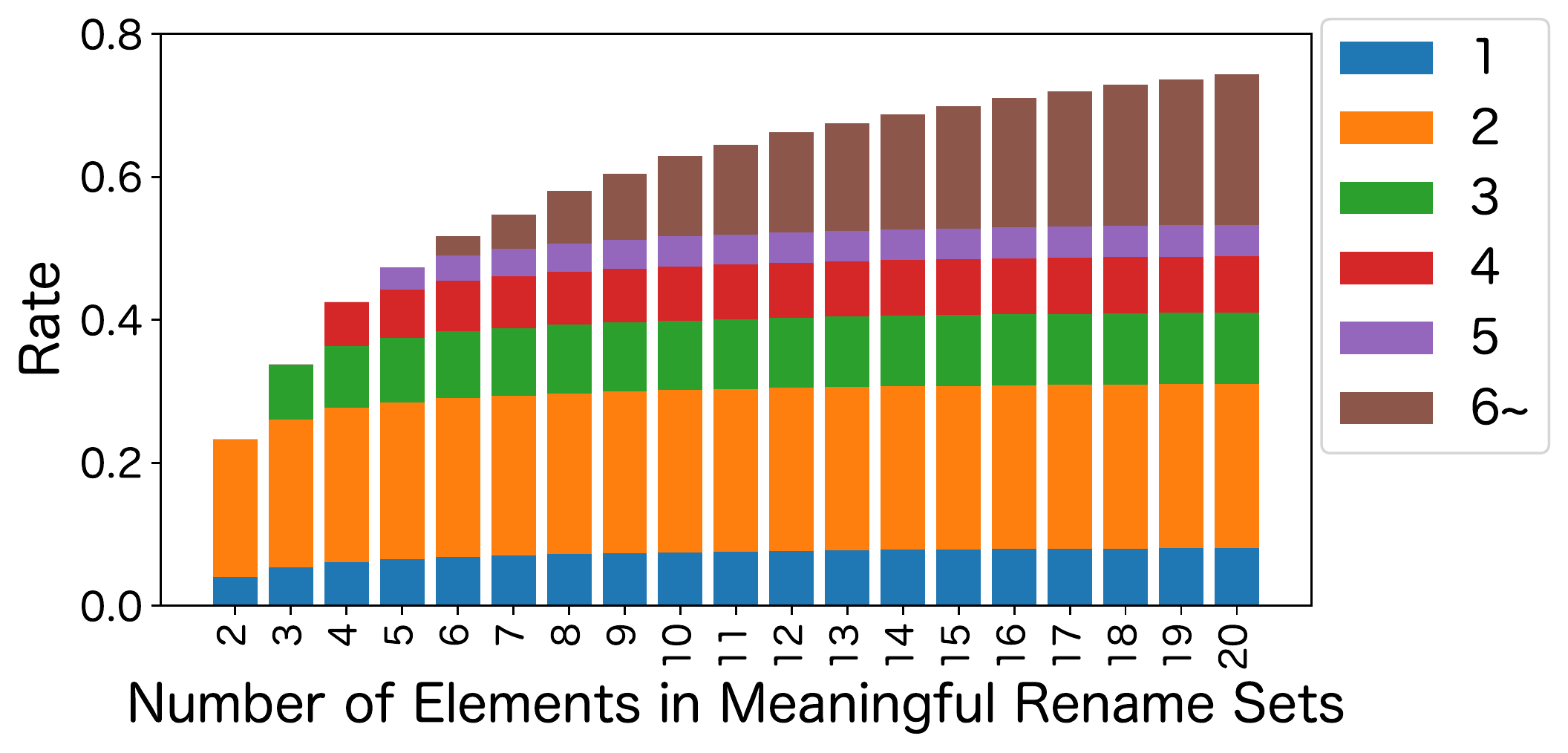}
  \caption{Cumulative distribution of the rate of co-renamings, colored by the number of unique identifier names.}\label{f:rq1_unit_size_unique}
\end{figure}

\Figref{f:rq1_unit_size_unique} shows a cumulative distribution of the percentage of the total number of elements of the meaningful rename sets for each number of the elements of the meaningful rename sets, colored by the number of unique identifier names.
The vertical axis represents the rate of meaningful rename sets, whereas the horizontal axis represents the maximum number of elements of the meaningful rename sets.
For example, the blue area in the figure shows the rate of meaningful rename sets that have only one unique identifier name.

The number of elements of a co-renaming was small.
From \figref{f:rq1_unit_size_unique}, we can observe that meaningful rename sets with less than six elements account for more than 50\% of all the meaningful rename sets considered as co-renaming.
Furthermore, 70\% of all the meaningful rename sets had no more than 20 elements.

On the other hand, most of the co-renamings consisted of different identifier names.
\Figref{f:rq1_unit_size_unique} shows that the rate of co-renamings with less than five unique identifier names did not increase even if the number of elements in meaningful rename sets increased.
In other words, there were few identifiers in a meaningful rename set whose names were the same as those of other identifiers in the meaningful rename set.
This result indicates that when recommending co-renamings, it is essential to consider how identifier names were renamed, i.e., operational chunks, not only to recommend identifiers matching names with those of renamed identifiers.

\Conclusion{57\% of all the renamings were co-renamings.
The co-renamings were composed of relatively few renamings.
It is important to consider operational chunks when recommending co-renamings.}

\subsection{\RQ{2}: \RQtwo}

\Heading{Motivation}
Because it is unnecessary to rename all the other identifiers to which an operational chunk of a renaming can be applied, we reveal the relationships between identifiers that are likely to be co-renamed.
For co-renamings detected in each repository, we evaluated the relationship between the co-renamed identifiers.

\Heading{Study Design}
We determined the relationships between the co-renamed identifiers and evaluated the 176 repositories, same as \RQ{1}.
From these repositories, 653,194 meaningful rename sets were created.
Of these, we analyzed 138,250 meaningful rename sets that contained more than two elements and detected 732,390 relationships.

We also limited the detection of relationships to meaningful rename sets that contain at least one renaming for a specific type of identifiers: Class, Method, Attribute, Parameter, and Variable.
By limiting the analysis, we could detect relationships between these types of identifiers, and those that are likely to be co-renamed.
For example, if we limit the detection of relationships to a meaningful rename set that contains at least one renaming of a Class, we do not analyze the meaningful rename sets that do not contain renamings of Classes.

\begin{figure}[tb]\centering
  \includegraphics[scale=0.28]{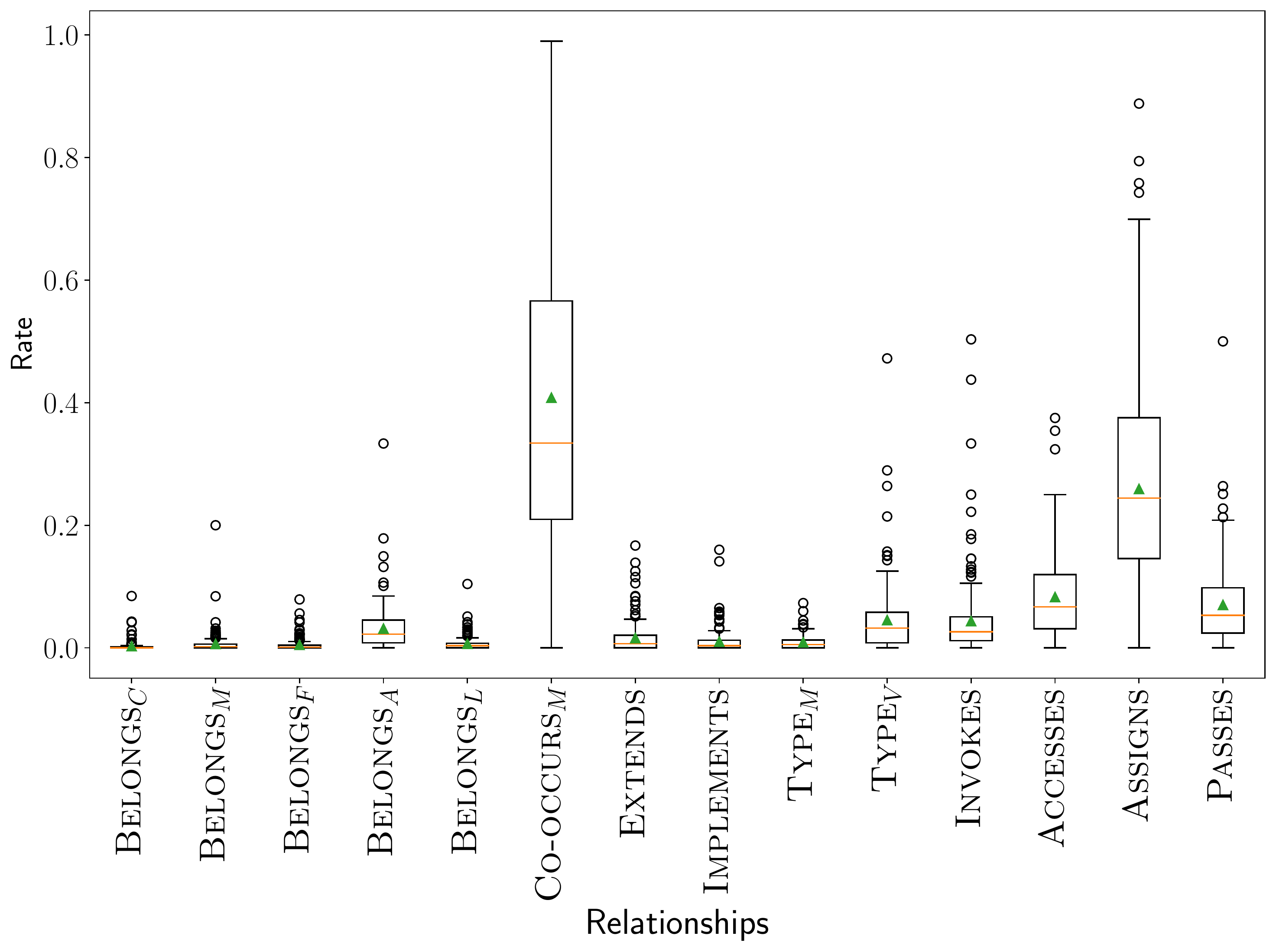}
  \caption{Rate of relationships.}\label{f:rq2_relation_all}
\end{figure}

\Heading{Results}
The most detected relationships were \CoOccursM and \Assigns.
\Figref{f:rq2_relation_all} shows a box plot of the rate of each relationship.
The vertical axis represents the rate of relationships, whereas the horizontal axis represents the kinds of relationships.
The mean values for the rate of \CoOccursM and \Assigns were 40.8\% and 25.9\%, respectively, the highest of all the relationships, while the other relationships were much lower.

\begin{figure*}[t]\centering
  \begin{minipage}{0.32\hsize}\centering
    \includegraphics[scale=0.17]{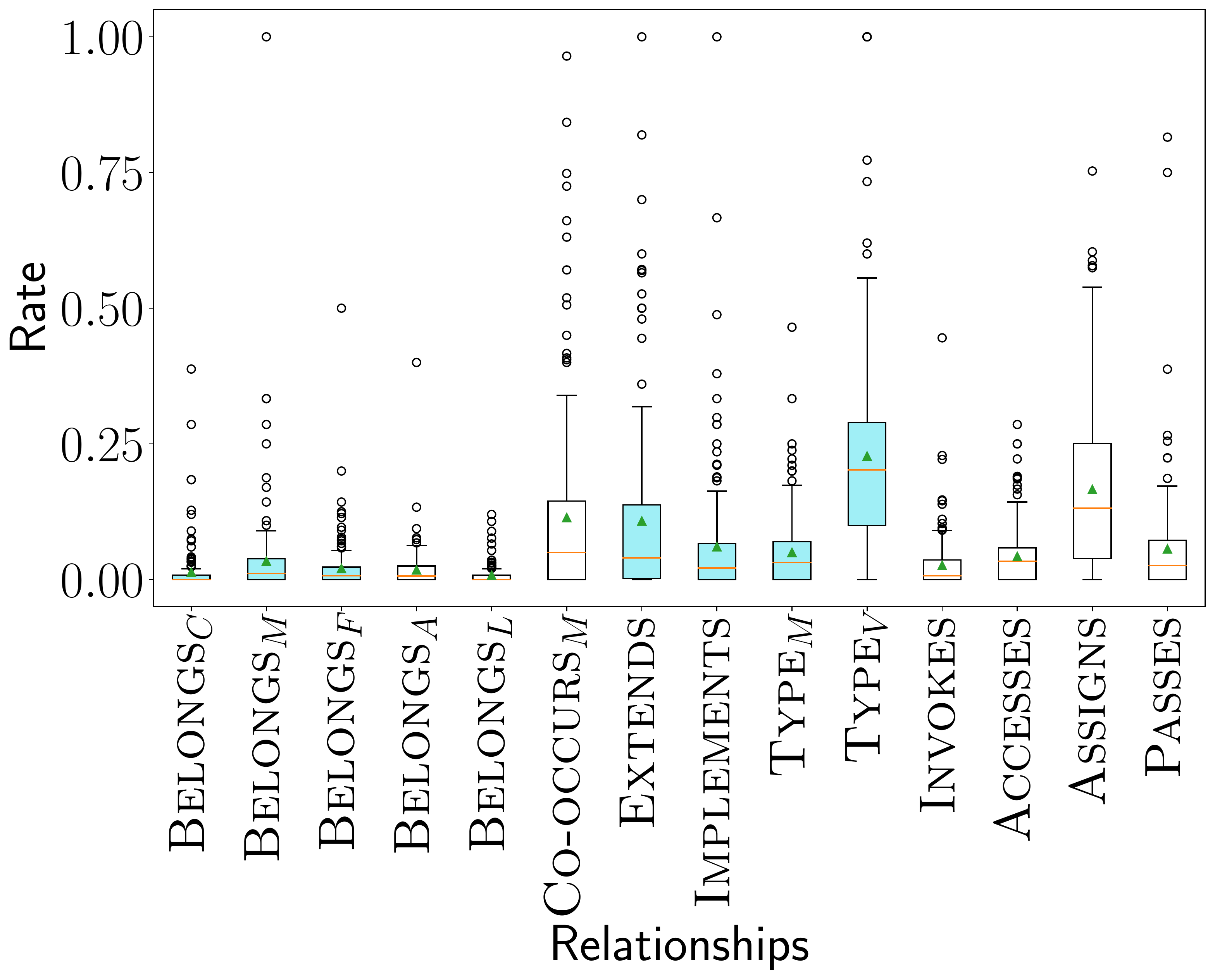}
    \subcaption{Classes.}\label{f:rq2_relation_class}
  \end{minipage}
  \begin{minipage}{0.32\hsize}\centering
    \includegraphics[scale=0.17]{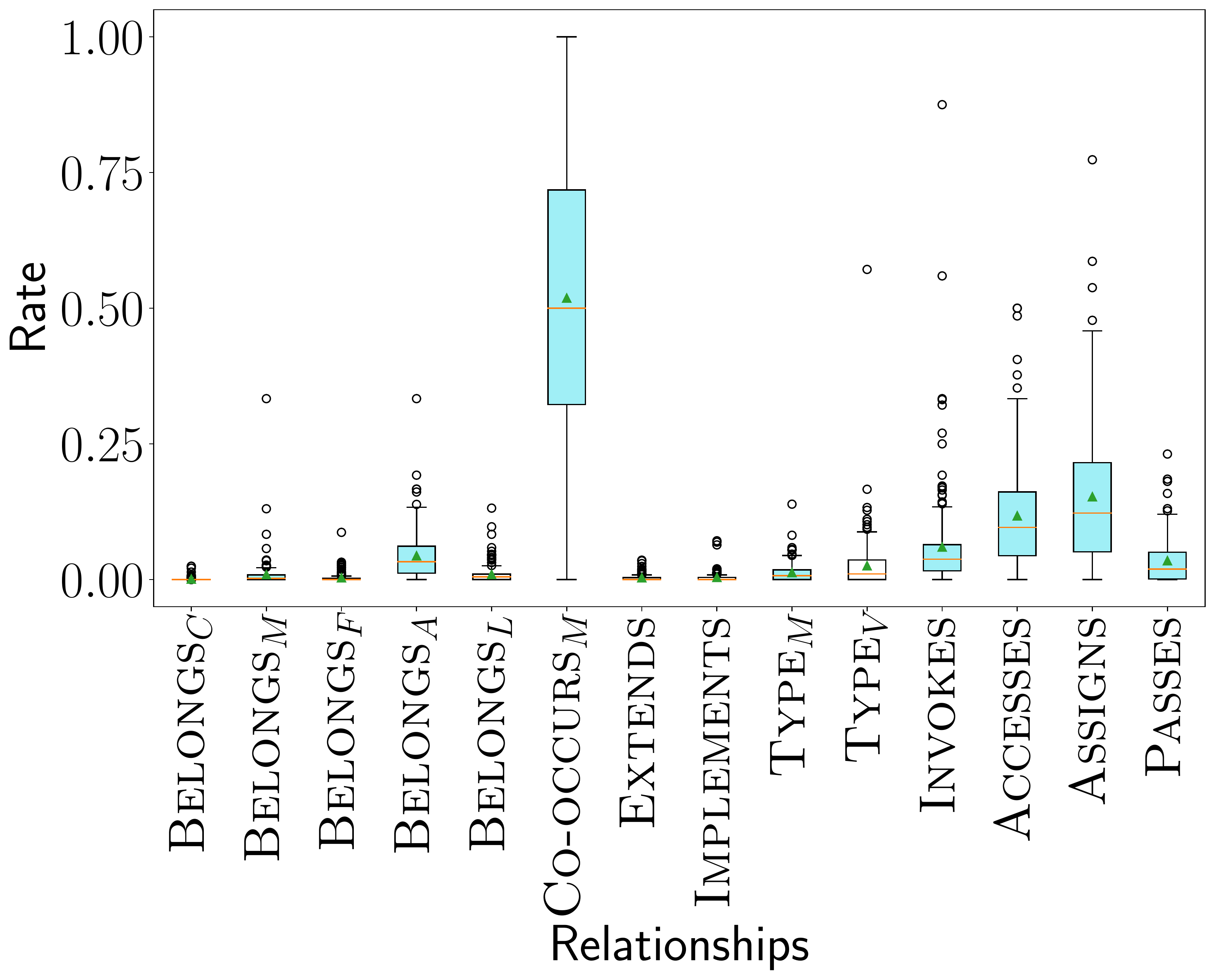}
    \subcaption{Methods.}\label{f:rq2_relation_method}
  \end{minipage}
  \begin{minipage}{0.32\hsize}\centering
    \includegraphics[scale=0.17]{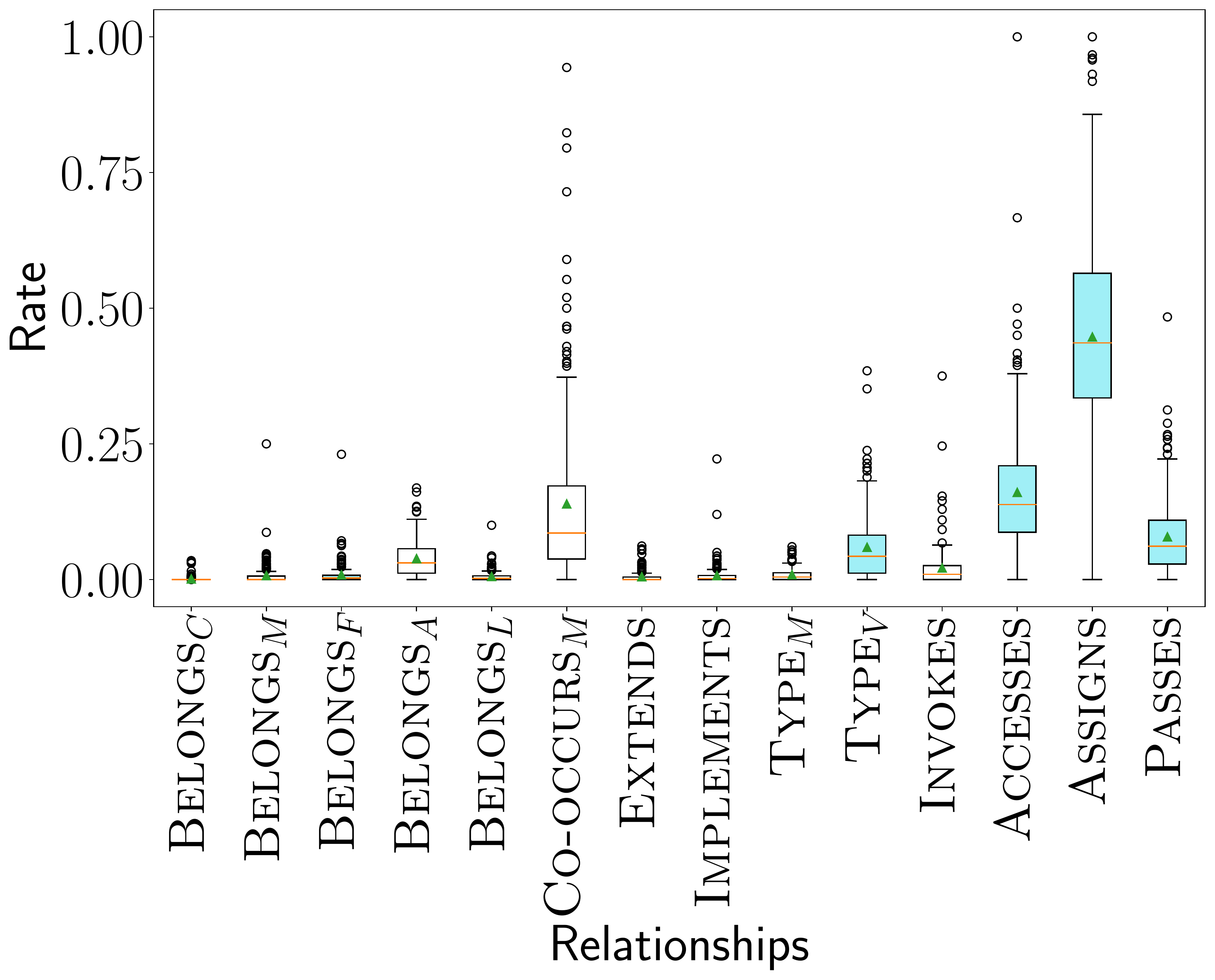}
    \subcaption{Attributes.}\label{f:rq2_relation_attribute}
  \end{minipage} \\\vspace{1.5em}
  \begin{minipage}{0.32\hsize}\centering
    \includegraphics[scale=0.17]{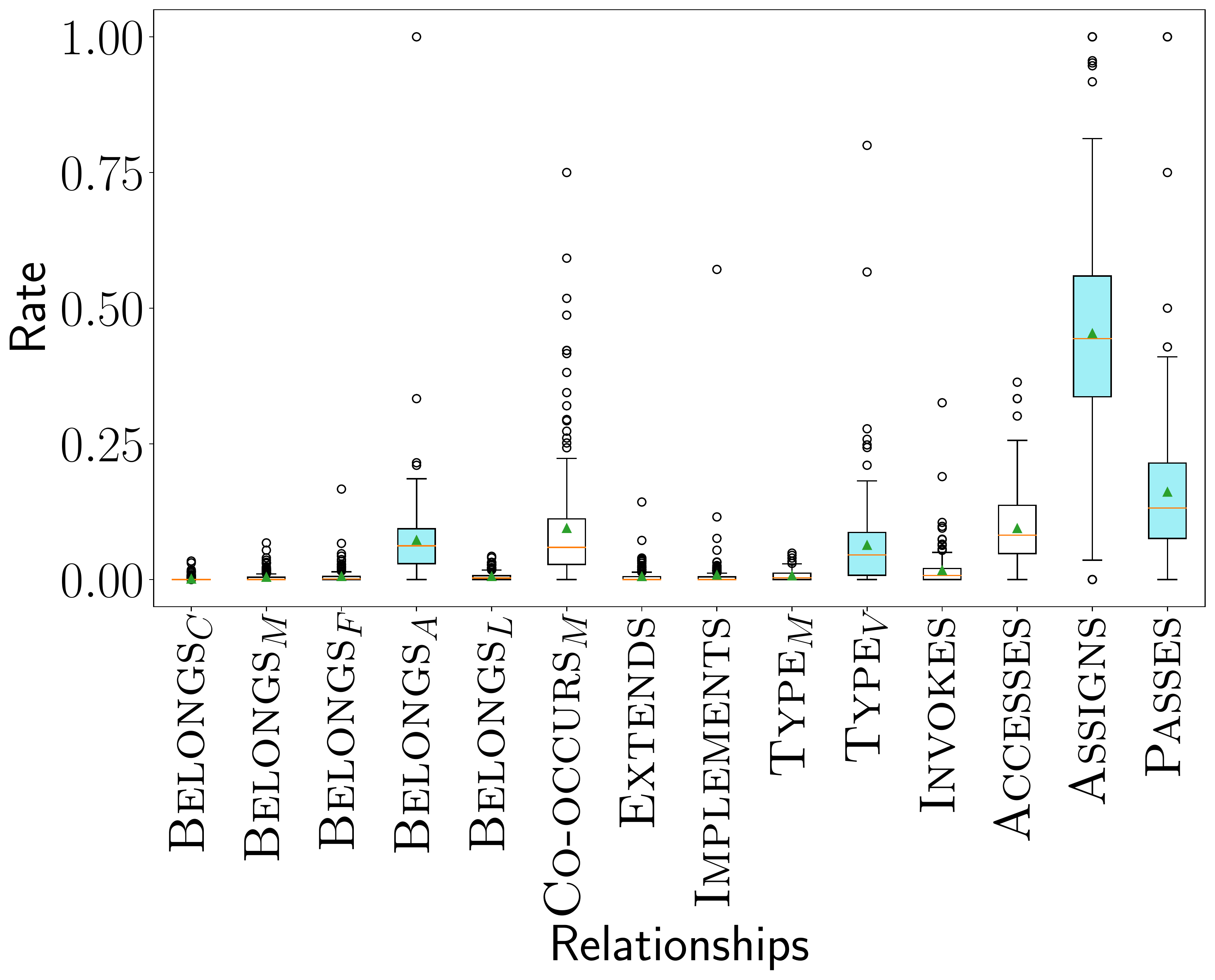}
    \subcaption{Parameters.}\label{f:rq2_relation_parameter}
  \end{minipage}
  \begin{minipage}{0.32\hsize}\centering
    \includegraphics[scale=0.17]{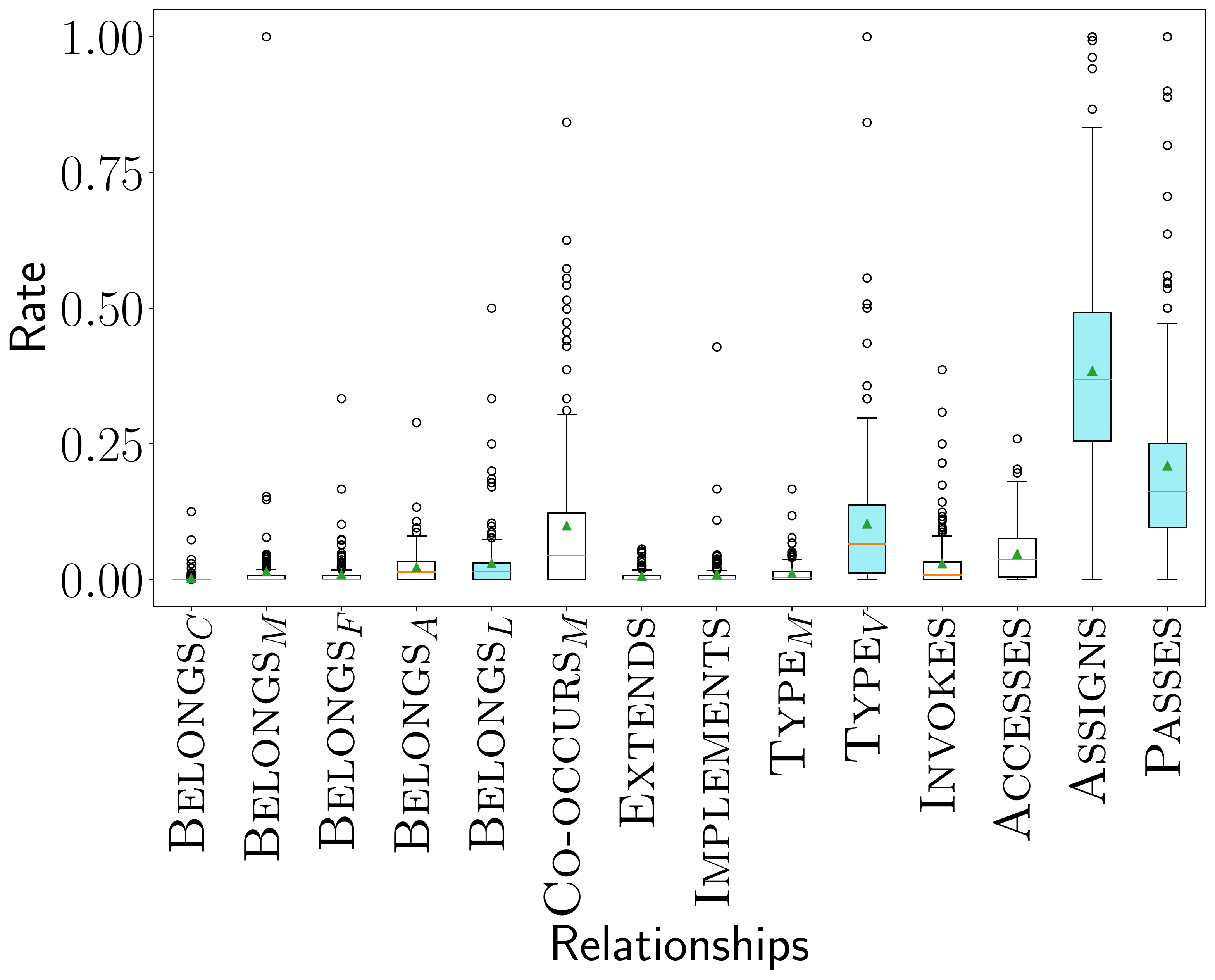}
    \subcaption{Variables.}\label{f:rq2_relation_variable}
  \end{minipage}
  \caption{Rate of relationships detected in co-renamings containing renamings of specific identifiers.}\label{f:rq2_relation_type}
\end{figure*}

\Figref{f:rq2_relation_type} shows a box plot of the rate of each relationship, limiting the analysis to co-renamings containing a specific type.
In each figure, the vertical axis represents the rate of relationships, whereas the horizontal axis represents the kinds of relationships.
Each box plot shows the distribution of particular relationships detected to the total relationships detected for each repository.
Light blue-colored ones are for the relationships that involve limited types of identifiers.
For example, \figref{f:rq2_relation_class} shows the rate of relationships when limiting the analysis to co-renamings that contain at least one renaming of Class, and the box plots of relationships that involve class renamings, i.e., \BelongsC, \BelongsM, \BelongsF, \Extends, \Implements, \TypeM, and \TypeV are highlighted as light blue.

The different distributions for each type of identifier in \figref{f:rq2_relation_type} indicate that the relationships in which identifiers are likely to be co-renamed differ depending on the type of identifier renamed.
\CoOccursM is the most detected relationship for Method, \Assigns for the other types.
Note that \TypeV is frequently detected relationship in Class; \Accesses in Attribute, \Passes in Parameters and Variables.
Each of these relationships is associated with the corresponding identifier type.
This result means that identifiers in a relationship to a renamed identifier are likely to be co-renamed with the identifier.

\Conclusion{
The distribution of the relationships differed significantly among the types of identifiers.
The most detected relationship was \CoOccursM for Method, \Assigns for the other types.
In addition, \TypeV frequently appeared in Class; \Accesses in Attribute, \Passes in Parameter and Variable.
These results suggest that identifiers that are likely to be co-renamed depend on the type of identifiers.}

\subsection{\RQ{3}: \RQthree}

\Heading{Motivation}
In \RQ{1} and \RQ{2}, we did not ignore inflection.
Thus, operational chunks that should be the same in multiple renaming may become a different operational chunk in each renaming.
Therefore, we evaluated the rate of co-renamings and the tendency of the relationships between the renamed identifiers in both cases of ignoring inflection and not and clarifying the effect of the inflection.

\Heading{Study Design}
In order to take inflections into account, when detecting operational chunks (in Section~\ref{s:detect}), we ignored inflection.
After splitting identifier names into word sequences in Section~\ref{s:detect}, for each word, we lemmatized it to ignore inflection using WordNetLemmatizer \cite{bird2009natural}.

WordNetLemmatizer can convert word forms within a part of speech, such as the singular and plural forms of nouns and the present and past tenses of verbs, but cannot convert word forms across parts of speech. 
For example, the plural form of a noun such as \textit{queries} can be converted to its singular form \textit{query}, but the noun \textit{creator} cannot be converted to the verb \textit{create}.

Then, we evaluated the effect of inflection on the operational chunks of renamings.
We evaluated the 176 repositories used in \RQ{1}.
If we do not ignore inflection, the operational chunks of renaming change only the inflection \Intent{Replace} instead of \Intent{Other}.
For example, in a renaming \Rename{node}{nodes}, when ignoring inflection an operational chunk \Other{node} is detected, while not an operational chunk \Replace{node}{nodes} is detected.
To evaluate the difference in detecting such operational chunks, we defined a new operational chunk \Intent{Inflect}, which represents inflection.
\begin{itemize}
  \item \Intent{Inflect}(\textit{words}): Inflection of words.
  For example, a renaming of \Rename{instance}{instances} involves an operational chunk of \Inflect{instance}.
\end{itemize}
When we detected \Intent{Other} from renaming $r$, we compared the words in the identifiers before and after the renaming from the beginning in a case-insensitive manner and detected \Intent{Inflect} each word that differs.
We compared the rate of each operational chunk, including \Intent{Inflect}, ignoring and not ignoring inflection.

We also evaluated the effect of inflection on the co-renamings in the same 176 repositories as \RQ{1} and \RQ{2}.
We evaluated the difference in the ratio of the total number of co-renamings to the total number of renamings and that in the number of meaningful rename sets $|\mathbb{U}|$ between the two cases of ignoring and not ignoring inflection for each repository.

Additionally, We evaluated the effect of the relationships between co-renamed identifiers in the same 176 repositories as in \RQ{2}.
For each repository, we created a set of meaningful rename sets $\mathbb{U} = \{U_1,\ U_2,\ \ldots \}$ ignoring inflection and a set of meaningful rename sets $\mathbb{U}^{\mathit{skip}} = \{U_1^{\mathit{skip}},\ U_2^{\mathit{skip}},\ \ldots \}$ without ignoring inflection.
We analyzed relationships between identifiers in the newly created meaningful rename sets, ignoring inflection, i.e., the meaningful rename sets included in a relative complement of $\mathbb{U}^{\textit{skip}}$ in $\mathbb{U}$ $(\mathbb{U} \setminus \mathbb{U}^{\textit{skip}})$.
The total number of newly created meaningful rename sets in all repositories was 119,165.

\Heading{Results}

\begin{figure}[tb]\centering
  \includegraphics[scale=0.35]{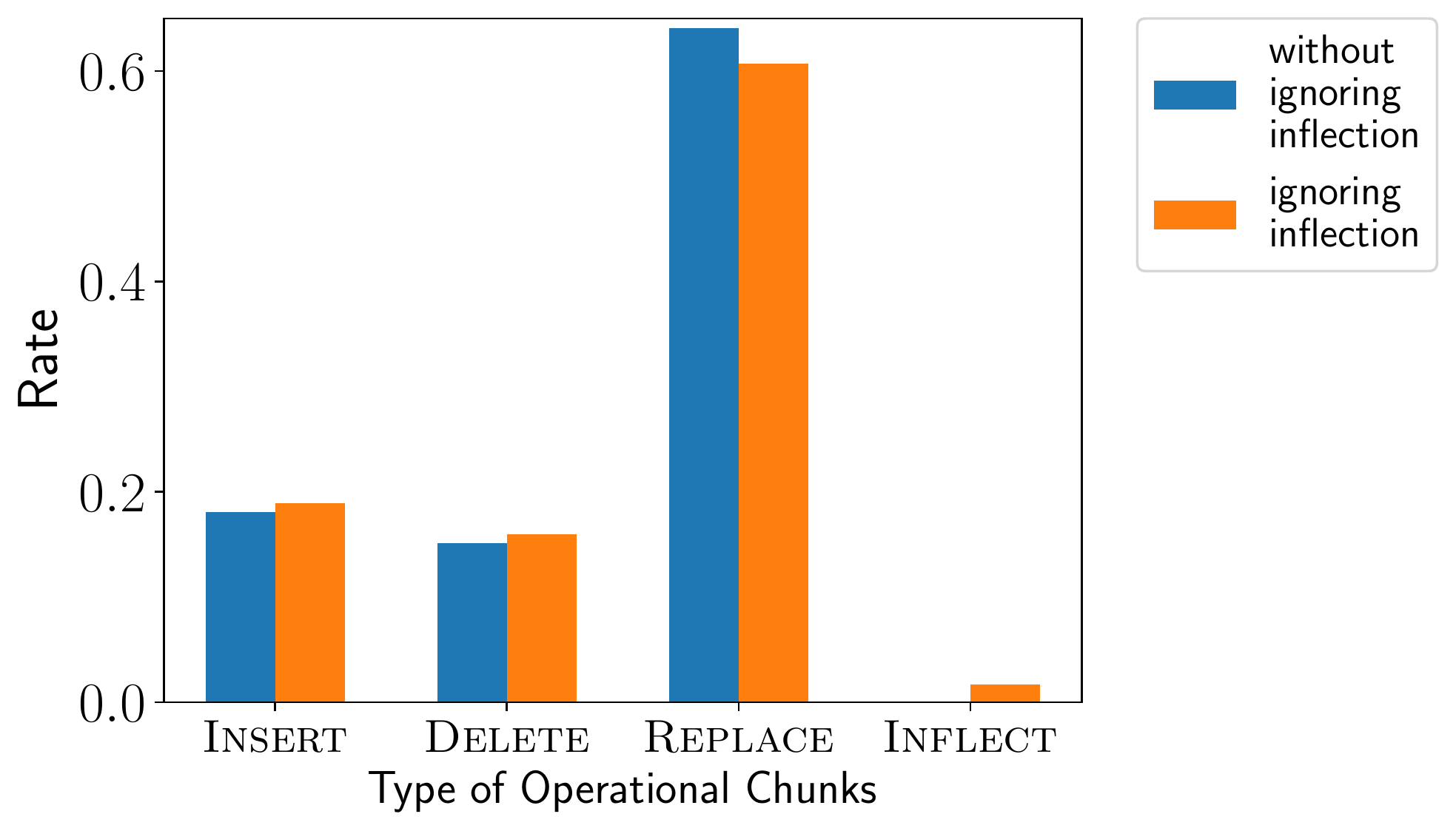}
  \caption{Rate of detected operational chunks.}\label{f:rq3_diff_rate}
\end{figure}

\Figref{f:rq3_diff_rate} shows the results of the evaluation of differences in operational chunks with and without ignoring inflection.
\Figref{f:rq3_diff_rate} shows the rate of operational chunks detected for each case with and without ignoring inflection in all repositories.
The vertical axis shows the rate of the total number of operational chunks, and the horizontal axis shows the types of operational chunks.
We excluded \Intent{Other} from the graph to focus on whether operational chunks changed due to inflection.
The rate of \Intent{Other} was 0.03 in both cases.

The rate of the number of operational chunks did not change significantly owing to the effect of inflection.
\Figref{f:rq3_diff_rate} shows that when we ignored inflection, the rate of \Intent{Replace} decreased, whereas the rate of \Intent{Inflect} increased.
One of the reasons for this result is that the operational chunk \Intent{Replace}, detected without ignoring inflection, changes to the operational chunk \Intent{Inflect} due to ignoring inflection.
However, the decrease in the rate of \Intent{Replace} was slight and did not significantly change the overall distribution.

The effect of inflection on the ratio of co-renamings to the total renaming was negligible.
The right side of \figref{f:rq1_rq3_co_rename} shows the rate of co-renamings detected by ignoring inflection.
The distribution, whose mean value of the rate of co-renamings detected with ignoring inflection is 0.58, is almost the same as that without ignoring inflection, as shown in the left side of \figref{f:rq1_rq3_co_rename} used in answering \RQ{1}.
The total number of meaningful rename sets when ignoring inflection was 651,389, whereas that when not ignoring inflection was 653,194, with almost no difference between the two cases.
However, there were cases where the number of detected operational chunks changed owing to ignoring inflection, and the number of meaningful rename sets also changed.
Because a meaningful rename set is created for each operational chunk, renaming that increases the number of operational chunks by ignoring inflection may create a new meaningful rename set and affect the results.

\begin{figure}[tb]\centering
  \includegraphics[scale=0.28]{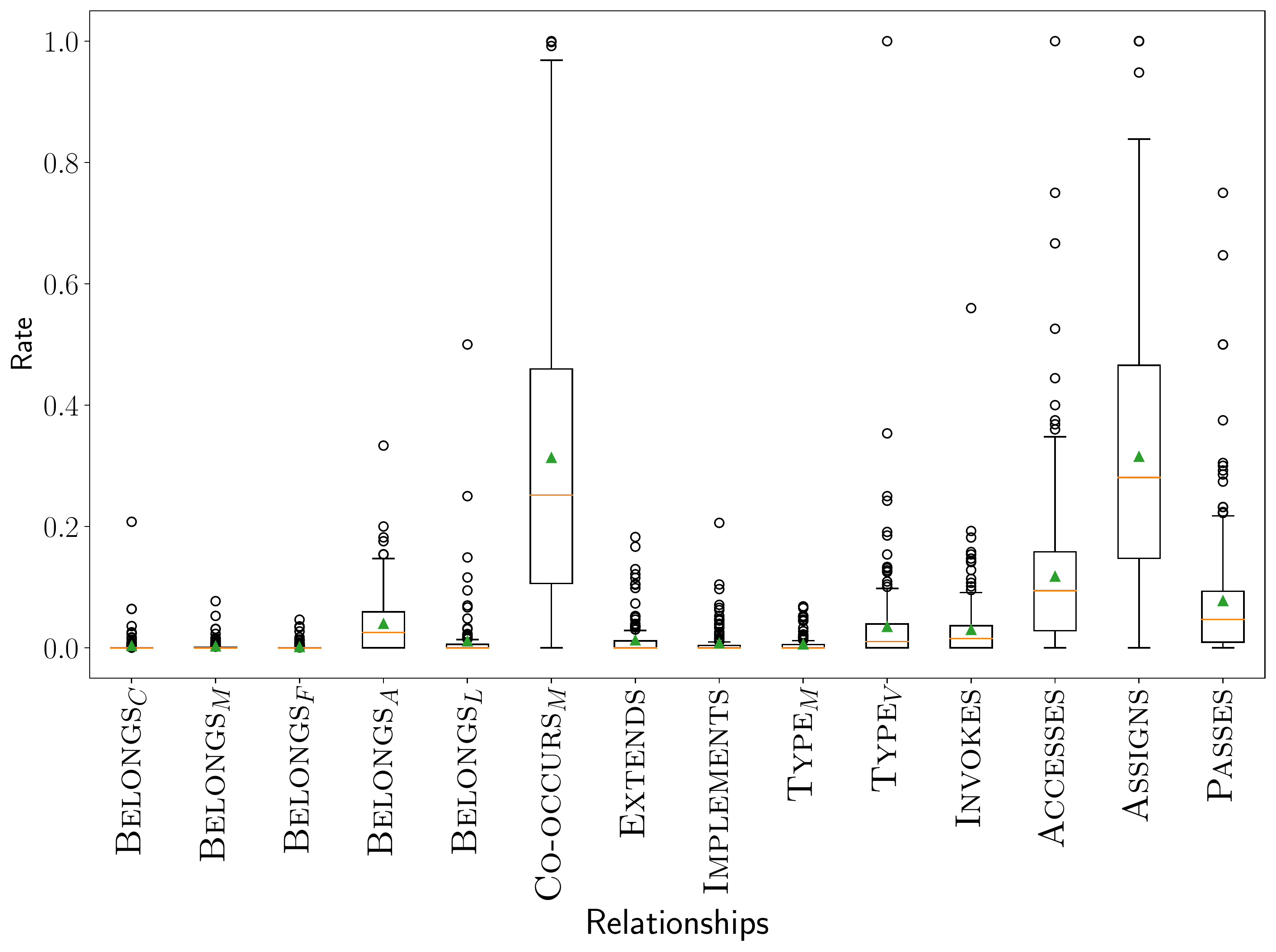}
  \caption{Rate of relationships in co-renamings affected by inflection.}\label{f:rq3_after_relation}
\end{figure}
\begin{figure*}[t]\centering
  \begin{minipage}{0.32\hsize}\centering
    \includegraphics[scale=0.17]{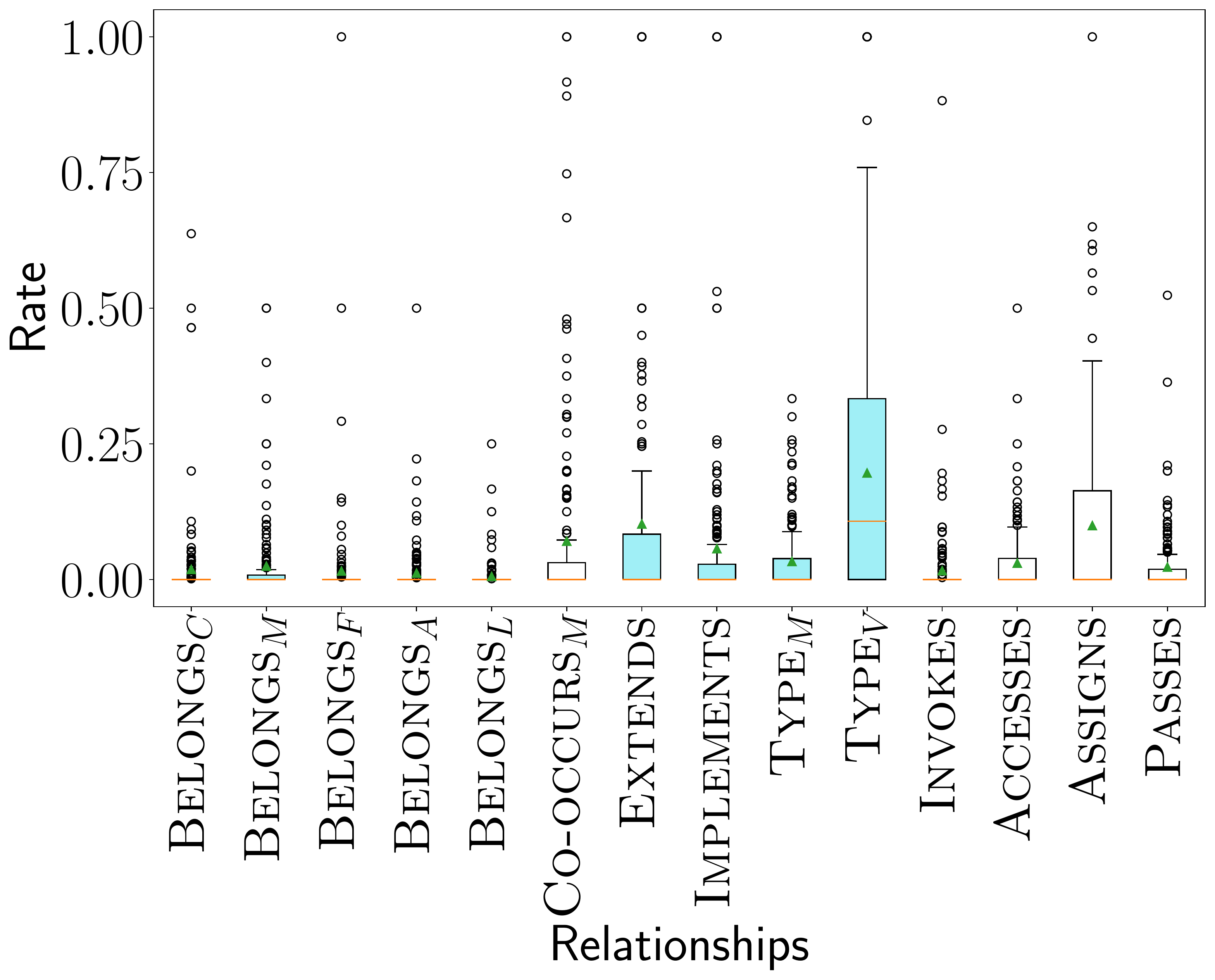}
    \subcaption{Classes.}\label{f:rq3_after_relation_class}
  \end{minipage}
  \begin{minipage}{0.32\hsize}\centering
    \includegraphics[scale=0.17]{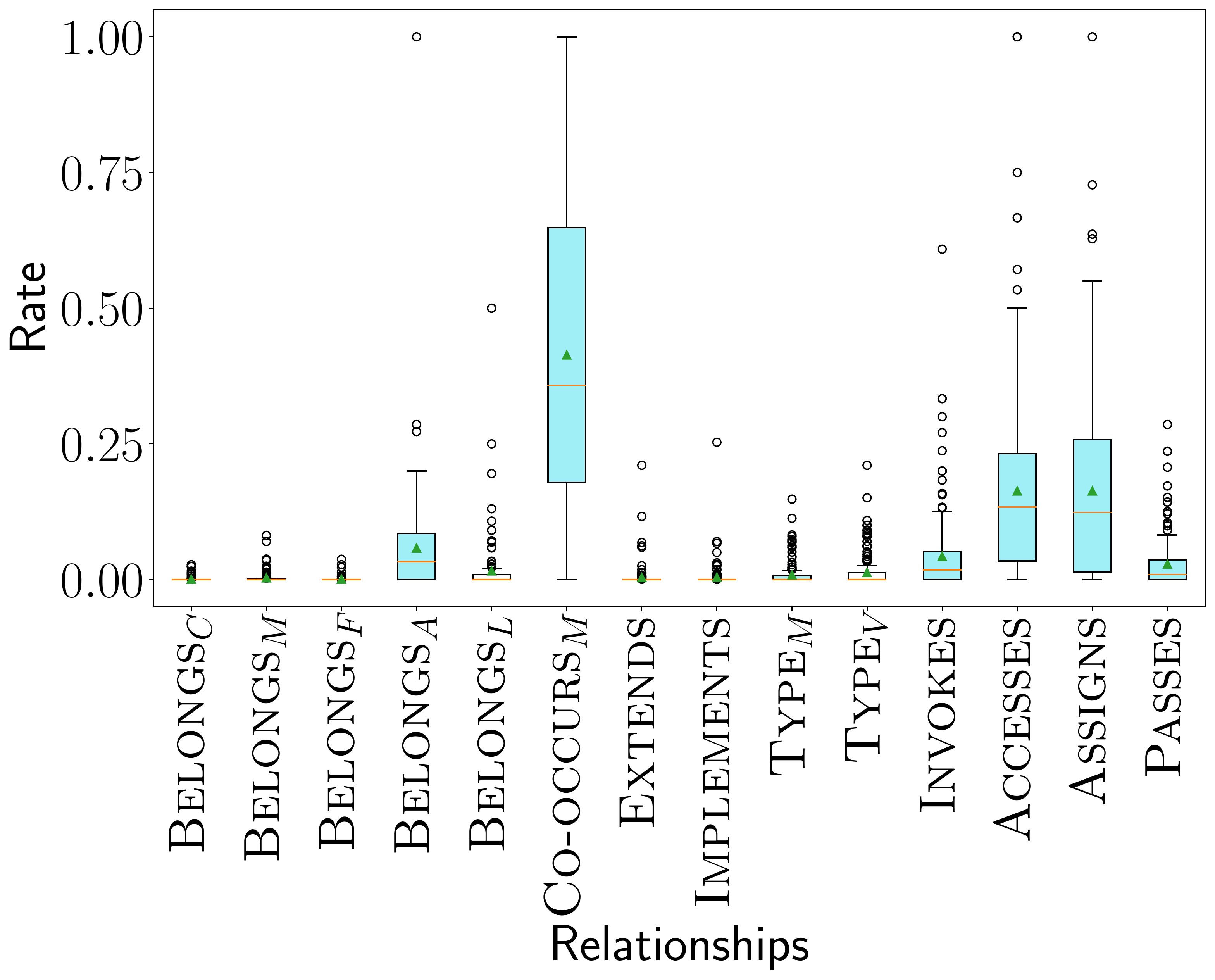}
    \subcaption{Methods.}\label{f:rq3_after_relation_method}
  \end{minipage} 
  \begin{minipage}{0.32\hsize}\centering
    \includegraphics[scale=0.17]{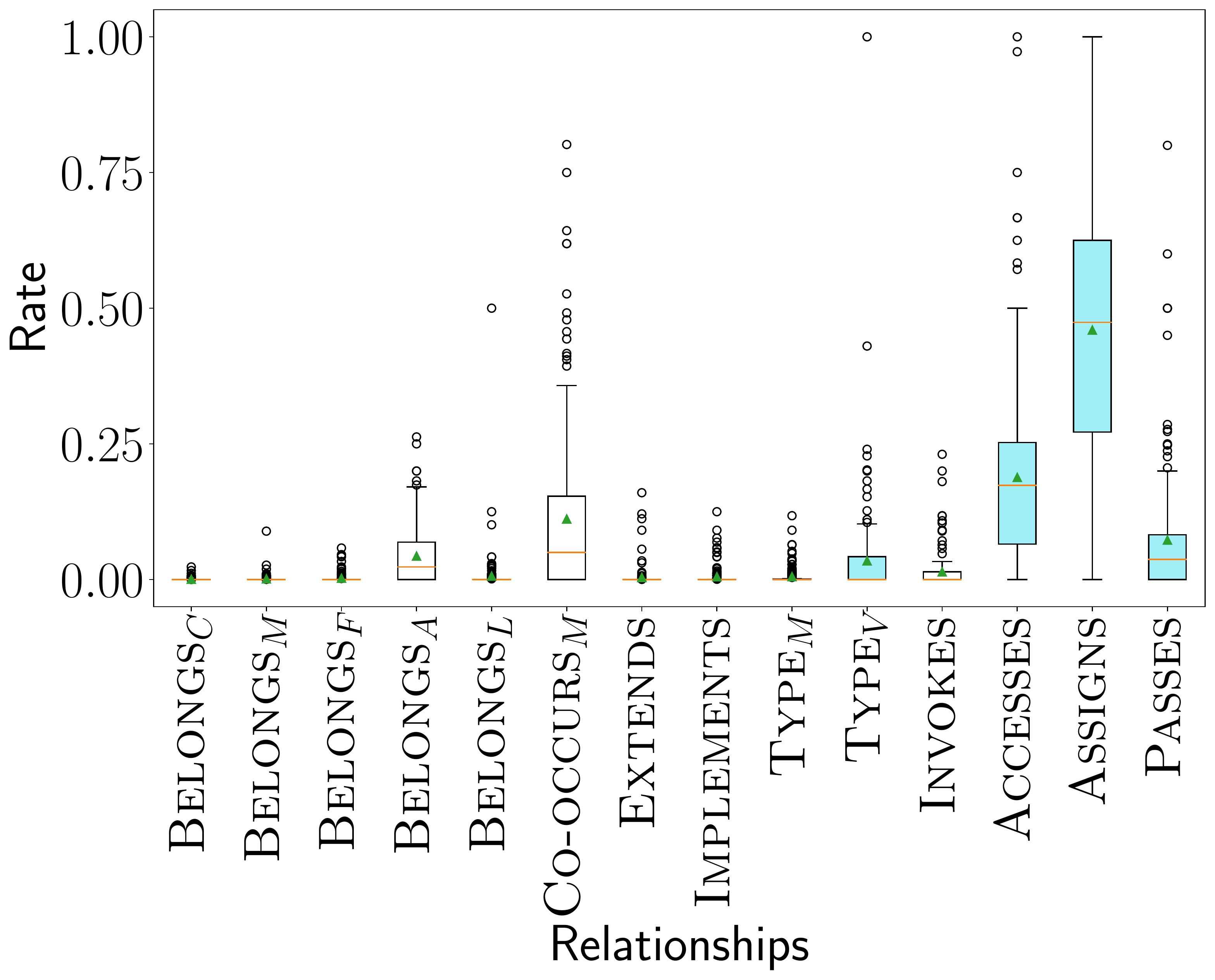}
    \subcaption{Attributes.}\label{f:rq3_after_relation_attribute}
  \end{minipage} \\\vspace{1.5em}
  \begin{minipage}{0.32\hsize}\centering
    \includegraphics[scale=0.17]{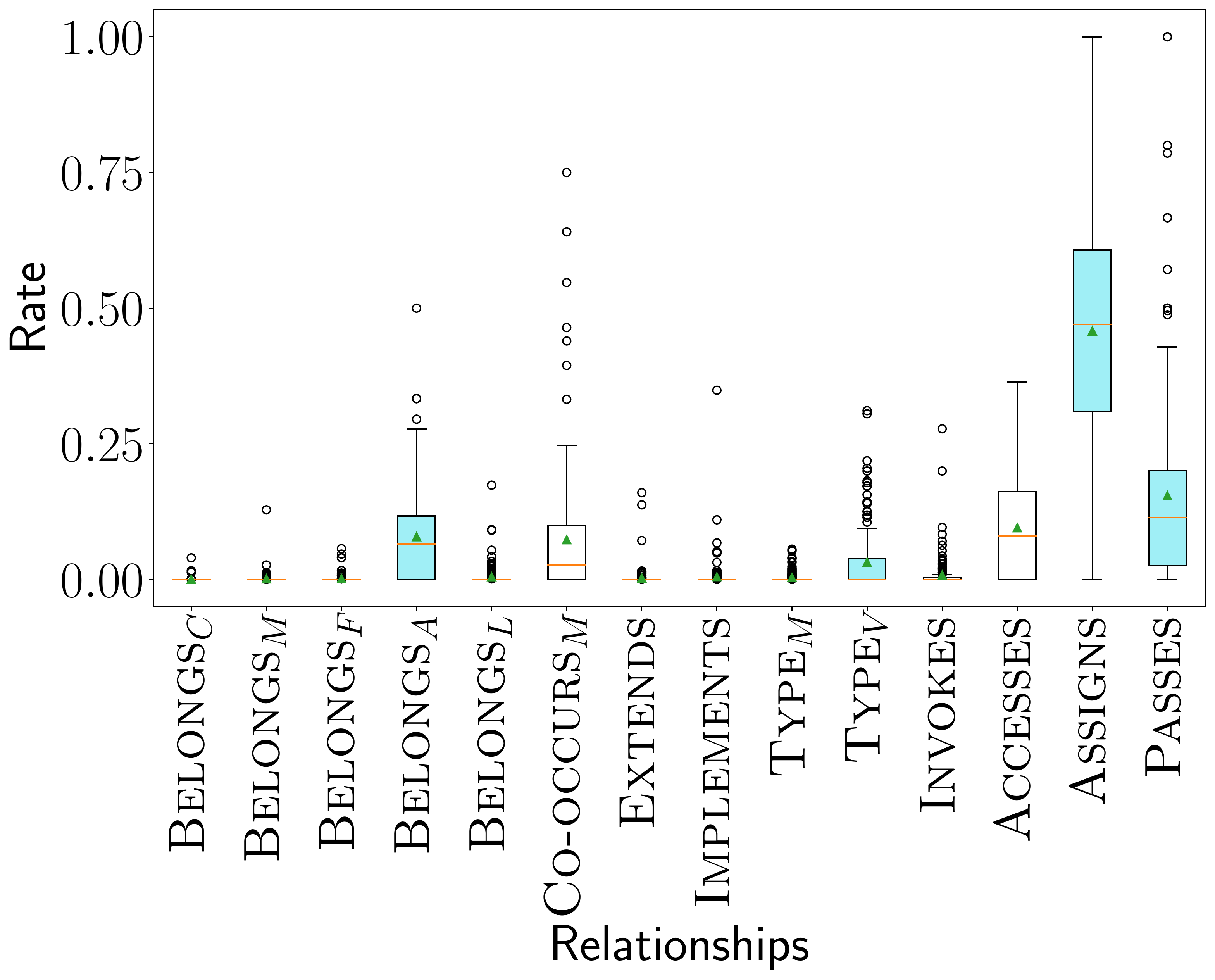}
    \subcaption{Parameters.}\label{f:rq3_after_relation_parameter}
  \end{minipage}
  \begin{minipage}{0.32\linewidth}\centering
    \includegraphics[scale=0.17]{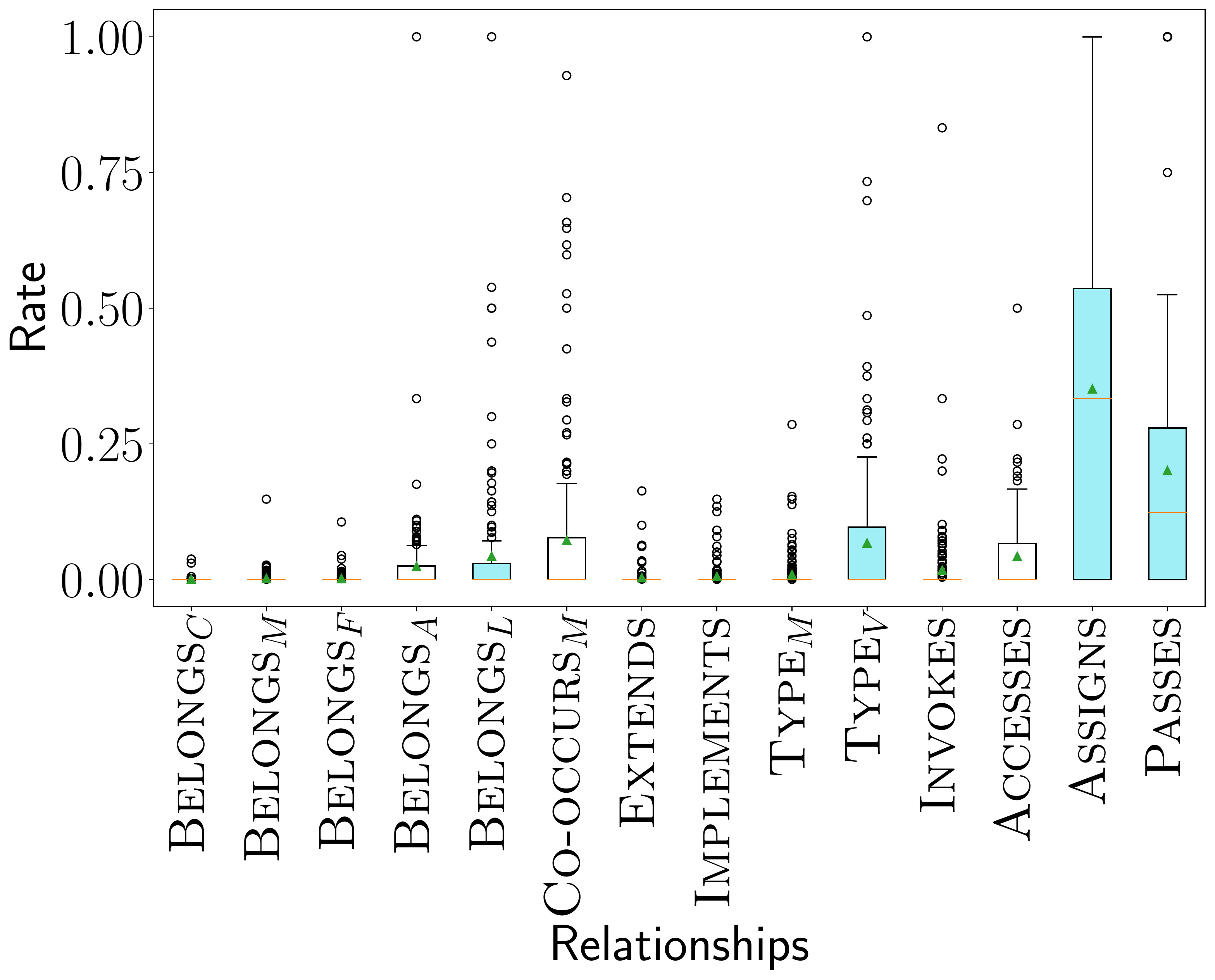}
    \subcaption{Variables.}\label{f:rq3_after_relation_variable}
  \end{minipage}
  \caption{Rate of detected relationships in co-renamings affected by inflection and containing renamings of specific identifiers.}\label{f:rq3_after_relation_type}
\end{figure*}

Figures~\ref{f:rq3_after_relation} and \ref{f:rq3_after_relation_type} show the results of analyzing relationships for the co-renamings affected by inflection.
\Figref{f:rq3_after_relation} is a box plot of the results of the rate of each relationship for each repository, as in \figref{f:rq2_relation_all}; \figref{f:rq3_after_relation_type} is a box plot of the results of the rate of each relationship obtained by limiting the detection of relations to co-renamings containing renamings of the specific type of identifiers for each repository, as in \figref{f:rq2_relation_type}.

The rate of relationships did not change significantly in general, except when analyzing co-renamings containing renamings of Class.
Comparing \figref{f:rq3_after_relation} with \figref{f:rq2_relation_all}, the rate of each relationship somewhat changed but not considerably different.
Comparing \figref{f:rq3_after_relation_type} with \figref{f:rq2_relation_type}, the rate of each relationship is also not considerably different for Method, Parameter, and Variable; however, for Class, the rate of \TypeV increases, while many other rates decrease.
This result may be due to ignoring the word inflection of the identifier name.
For example, if the instances \textit{query} and \textit{queries} of the class \textit{Query} are co-renamed to \textit{entry}, \textit{entries}, and \textit{Entry}, respectively, the number of \TypeV increases because of ignoring the inflection of the identifier name.
We assume that the rate of \TypeV increased owing to such co-renamings.

\Conclusion{There was a slight effect of inflection on the rate of operational chunks.
It also had no significant effect on the rate of co-renamings.
There was no significant effect in terms of relationships in general, except for an increase in the rate of \TypeV in the co-renamings containing renamings of Class.}

\subsection{Discussion}
In recommending co-renamings, we should change the identifiers recommended preferentially based on the type of renamed identifier and relationships between them.
In \RQ{2}, we found that the tendency of relationships between renamed identifiers was different for each type of renamed identifier.
We inferred that we can improve the accuracy of recommending co-renamings by recommending identifiers that have a relationship, which are likely to be co-renamed with a renamed identifier, based on the type of renamed identifier. 
For example, suppose that a class \Id{Sample} has methods \Id{addItem} and \Id{removeItem}, and \Id{addItem} is renamed to \Id{addElement}.
This renaming is for a method, and it involves \Replace{item}{element}.
In answering \RQ{2}, we found that the most detected relationship is \CoOccursM in method renamings.
Therefore, we can recommend renaming identifiers in a \CoOccursM relationship with \Id{addItem}, i.e., \Id{removeItem} to \Id{removeElement}.
We found a case similar to this example in neo4j\footnote{https://github.com/neo4j/neo4j/commit/f24af6d}.
In this project, there were renamings including \Delete{count} such as \Rename{countPins}{pins} and \Rename{countBytesRead}{bytesRead}.
These renamings were only for methods; no renamings occurred for other type identifiers including \Word{count} like an attribute \Id{pageCount}.
Changes in interface methods probably caused this, so it was not necessary to rename all the identifiers, including \Word{count}.

However, the effect of inflection is negligible, and we should pay little attention to the inflection.
In \RQ{3}, we evaluated the changes in the rate of operational chunks and co-renamings due to ignoring inflection and the trend in relationships between co-renamed identifiers in meaningful rename sets affected by ignoring inflection. 
However, the results of \RQ{3} did not significantly differ from the results of \RQ{1} and \RQ{2}.
This result suggests that we do not need to pay attention to inflection in recommending co-renamings.
Although the detection rate of \TypeV increased for the co-renamings containing renamings of Class when ignoring inflection, this relation also had a high detection rate when not ignoring inflection.
Therefore, inflection does not seem to affect the recommendation of co-renamings.
However, in some cases, the number and type of detected operative chunks changed depending on whether we ignored inflection or not.
Thus, it is necessary to evaluate the case such that the number of operational chunks changes because of ignoring inflection in detail.

Some relationships may have a direction of propagation of renamings, and it is necessary to evaluate the directions of the relationships.
We can consider the direction of a relationship when one renaming of two renamings in a relationship causes the other renaming.
In \RQ{3}, we detected many \Passes for co-renamings containing renamings of Parameters or renamings of Variables.
Liu \etal \cite{liu2016nomen} evaluated identifiers with \Passes relationships and found that in many cases, it is better to rename arguments (i.e., Variables) than Parameters. 
Based on Liu \etal and our results, we deduced that most of the co-renamings in the \Passes relationship are due to renamings of  Parameters, which causes renamings of Variables.
By evaluating the directions of relationships, we may discover more valuable things for a recommendation of co-renamings.

\subsection{Threats to Validity}\label{s:threat}

\Heading{Internal Validity}
The tools used in this study may have influenced the results.
We used RefactoringMiner\cite{tsantalis2018accurate} to extract renamings; but it is not the latest version, but the October 2018 version.
However, the latest version of RefactoringMiner might yield different results from our evaluation.
In addition, we used WordNetLemmatizer\cite{bird2009natural} to ignore inflection, but we did not care about the accuracy of the tool.

The relationships between identifiers in our evaluation do not cover all the relationships between the identifiers that tend to be co-renamed.
In fact, in several meaningful rename sets, we could not detect any relationships.
We can evaluate the relationships in more detail by examining these meaningful rename sets and defining new relationships.

\Heading{External Validity}
The results of our evaluation may not be the same for all repositories in general.
We evaluated only open-source Java repositories.
In repositories we did not evaluate, the trend of co-renamings and the degree of the influence of inflection might be different from our results.
We evaluated 176 repositories of various sizes to address this issue: from 2,482 to 1,907,462 LOC and from 131 to 255,628 commits.

\section{Related Work}\label{s:relatedwork}
      
Identifiers in a program have an important role in program understanding \cite{madani-csmr2010,caprile-icsm2000,lawrie-wcre2010}.
Deissenboeck and Pizka stated that identifiers account for approximately 70\% of program texts \cite{deissenboeck-sqj2006}.
Corazza \etal reported that developers could smoothly communicate with each other by giving appropriate identifier names that reflect the developer intention and the domain knowledge \cite{LINSEN-corazza-icsm2012}.
If identifiers are named appropriately, developers can infer their intention and behavior \cite{RefactoringWorkbook}.

Peruma examined identifier renamings along with commit messages, their data types, and refactorings before and after them and reported that developers tended to rename identifiers to narrow their meanings, and 17.39\% of all the renamings changed their corresponding data types \cite{peruma2020contextualizing}.

Arnaoudova \etal proposed REPENT \cite{REPENT-arnaoudova-tse2014}, an approach to automatically detect and classify identifier renamings in source code. Based on a natural language processing technique, REPENT classifies renamings into different categories such as meaning-preserved, narrowed, or broadened.
Peruma \etal presented an empirical study of how identifiers were renamed, with an attention of whether the meaning of identifiers were to be narrowed or broadened \cite{how-and-why}.
Their classifications are more semantic-oriented, whereas our types of operational chunks are more lexical, based on the sequence of words in identifiers.
In addition, REPENT and the study by Peruma \etal is focused on the classification of renaming instances, whereas our analysis focused on the co-renamings, i.e., the relationship between renamed identifiers.

Several studies tried to perform a parsing for investigating the structure of identifiers with a specialized grammar and capturing a better meaning of the word fragments in identifiers \cite{NEWMAN2020110740,8919135}.
In addition, abbreviations are common in identifiers, and the expansion of such abbreviations can reveal the meaning of identifiers and relationships among them \cite{hill2008amap,8454758,8377888}.
These approaches can be more reliable than a simple inflection detection in our approach, and an extension of the inflection analysis in this paper may have room for improvements by embedding such sophisticated analyses approaches into our study framework.

Techniques to correct identifier names have been proposed for avoiding inconsistent identifier names that hinder program understanding.
Some of them normalize identifiers so that the naming conventions are consistent for the entire program.
Caprile and Tonella proposed a technique to generate a new identifier name by normalization using pre-defined rules and dictionaries \cite{caprile-icsm2000}.
Surafel \etal proposed a technique to suggest identifier names using an ontology generated from source code \cite{surafel-CSMR2013}.
Kashiwabara \etal proposed a method to recommend an appropriate verb to be used as a method name using association rule mining \cite{kashiwabara-CSMR2014}.
Lawrie \etal proposed a technique to calculate the similarity of vocabulary in identifiers using information retrieval technology and normalizing them \cite{lawrie-wcre2010}.

Meanwhile, the renaming by developers frequently occurs during software refactoring.
Because an inconsistent renaming leads to inconsistent identifier names, preventing such renaming is also important.
Fowler's refactoring catalog \cite{Refactoring} contains rename refactorings such as Rename Method.
Renaming is the most used refactoring operation \cite{murphyhill-tse2012,murphy2006java}, and developers are able to rename a specific identifier using existing refactoring tools consistently.
However, these tools do not have a co-renaming feature of related identifiers.
Liu \etal \cite{Expanding-Conducted-Rename-Refactorings} proposed RenameExpander, which recommends the renamings of identifiers related to the identifier renamed by developers.
However, they did not investigate what relationships frequently occurred.
Thies and Roth proposed a rename refactoring recommendation approach based on the assignments for variables \cite{Thies}, which is similar to the relationship of \Assigns in our approach.
Our results can provide not only empirical evidence of the effectiveness of such approaches but also supportive opportunities to improve their recommendation results with a deep look at the type of relationships.

Multiple renamings tend to occur at once in software refactoring.
If there is an identifier related to another identifier to be renamed, it may be necessary to rename it too.
Saika \etal analyzed a refactoring operation history performed by developers and reported that a series of refactorings such as multiple renamings is often performed \cite{saika-iwesep2014}.
There are several attempt of an IDE support for applying a sequence of continued refactoring operations \cite{maruyama-icse2017}.
The results of our empirical study can support the design of such refactoring tools.
Programmer-friendly refactoring tools can be improved by being aware of the type of the identifier to be triggered to rename and providing tailored support according to what entities are to be renamed.

\section{Conclusion}\label{s:conclusion}
In this study, we evaluated the effects of co-renamed identifiers on the relationships between identifiers and on the word form changes of words in the identifiers.
The results showed that half of the identifier renamings occurred together with other renamings. Additionally, the relationships between identifiers that are likely to be co-renamed differ depending on the type of the renamed identifiers.
Finally, a slight effect was observed for word form changes.
These results suggest that, in recommending a renaming, it is beneficial to prioritize candidate identifiers to be co-renamed for each type of the renamed identifier.

Future work will include further definition and evaluation of relationships, and the evaluation of the direction of relationships.
In some co-renamings of identifiers, the relationships considered in this study were not detected at all.
Defining new relationships for these co-renamings may lead to discovering additional characteristics for recommending new identifiers to be renamed.
Additionally, there was a possibility that some of the relationships had a direction.
It is possible that more detailed characteristics of the relationships can be revealed by determining the identifier renamings that cause another, e.g., by considering the time series of the changes.

The dataset of co-renamed identifiers used in this study is publicly available \cite{datasets}.

\section*{Acknowledgments}
This work is partly supported by JSPS KAKENHI (JP22H03567, JP21H04877, JP21K18302, and JP21KK0179).


\end{document}